\documentclass[%
reprint,
superscriptaddress,
 amsmath,amssymb,
 aps,
prb,
]{revtex4-1}

\usepackage{graphicx}
\usepackage{dcolumn}
\usepackage{bm}
\usepackage[dvipdfmx,colorlinks,bookmarks=true,citecolor=blue,linkcolor=blue,urlcolor=blue, breaklinks=true]{hyperref}
\usepackage{mathrsfs}
\usepackage{epstopdf}
\usepackage{txfonts}
\usepackage{cases}
\usepackage{float}

\begin{document}

\preprint{APS/123-QED}

\title{Topological and dynamical properties of a generalized cluster model in one dimension}

\author{Takumi Ohta}
 \email{takumi@yukawa.kyoto-u.ac.jp}
\affiliation{%
 Yukawa Institute for Theoretical Physics, Kyoto University,
 Kyoto 606-8502, Japan
}%
\author{Shu Tanaka}%
\affiliation{%
 Waseda Institute for Advanced Study, Waseda University,
 Tokyo 169-8050, Japan
}%
\author{Ippei Danshita}
\affiliation{%
 Yukawa Institute for Theoretical Physics, Kyoto University,
 Kyoto 606-8502, Japan
}%
\author{Keisuke Totsuka}
\affiliation{%
 Yukawa Institute for Theoretical Physics, Kyoto University,
 Kyoto 606-8502, Japan
}%

\date{\today}

\begin{abstract}
We study the ground-state phase diagram and dynamics of the one-dimensional cluster model
with several competing interactions.  
Paying particular attention to the relation between the entanglement spectrum (ES) and 
the bulk topological (winding) number, 
we first map out the ground-state phases of the model 
and determine the universality classes of the transitions from the exact solution.  
We then investigate the dynamical properties during interaction sweeps through the critical points
of topological phase transitions.  
When the sweep speed is slow, the correlation functions and the entanglement entropy 
exhibit spatially periodic structures.
On top of this, the levels in the ES oscillate temporally during the dynamics.  
By explicitly calculating the above quantities for excited states, 
we attribute these behaviors to the Bogoliubov quasiparticles generated near the critical points.  
We also show that the ES reflects the strength of the Majorana correlation even for the excited states. 
\end{abstract}

\pacs{Valid PACS appear here}
\maketitle


\section{\label{sec:Introduction}Introduction}
The cluster model in one dimension has attracted attention in statistical physics,
condensed matter physics, and quantum-information science.\cite{Suzuki-1971,Raussendorf-2001,Raussendorf-2003,Skrovseth-2009,Son-2011,Smacchia-2011,Lahtinen-2015} 
In the early stage of its studies,  the one-dimensional cluster model and its variants were investigated 
as an example of a series of one-dimensional spin models that can be exactly solved by mapping to 
free fermions.\cite{Pfeuty-1970,Suzuki-1971,Suzuki-2013} 
Looked upon as fermion models, this class of models may be thought of as describing one-dimensional 
$p$-wave superconductors (Kitaev chain) with longer-range hopping and pairing.\cite{Kitaev-2001}    
The ground state of the cluster model, called the cluster state, shows topological properties, e.g., 
the existence of unpaired Majorana fermions appearing at the boundaries of the system. 

Toward the realization of quantum information processing, interest in the cluster model has revived.  
In the 2000s, one-way quantum computation\cite{Raussendorf-2001} and measurement-based quantum computation\cite{Raussendorf-2003} have been proposed.
The cluster state can be used to implement them since a highly entangled state must be prepared 
for the resource state.\cite{Fujii-2015}
Recently, there is extensive research on the cluster model such as the entanglement properties\cite{Skrovseth-2009,Smacchia-2011,Ohta-2015}
and the robustness of the cluster state against thermal excitations or randomness.\cite{Fujii-2013,Bahri-2015}
Entanglement carries information on the complexity of the quantum states,
which is related to the efficiency of the computation. 
In addition, the one-dimensional cluster model has a potential for realization in experiments of cold atoms on a zigzag ladder
by introducing three-spin exchange interaction.\cite{Pachos-2004}
For the quantum computation in an experimental setup,
it is important to know the stability of the cluster state and properties of entanglement.

Entanglement is also an important concept to study the quantum phase transitions and topological properties.
There are mainly two ways to quantify entanglement:
entanglement entropy (EE) and entanglement spectrum (ES).
The EE, which is defined as the von Neumann entropy of the reduced density matrix,
characterizes topological properties in the ground states.
\cite{Kitaev-2006,Levin-2006,Furukawa-2007,Katsura-2007,Katsura-2008,Katsura-2010,Jiang-2012}
The scaling of the EE has information on the topological field theory of critical points such as the central charge.
\cite{Holzhey-1994,Calabrese-2004}
The spectrum of the eigenvalues of the reduced density matrix, called the ES,
contains more information than the EE.\cite{Li-2008}
The ES reflects the criticality of the entanglement Hamiltonian derived from the reduced density matrix.\cite{Cirac-2011,Lou-2011,Tanaka-2012}
In addition, the level structure of the ES
has a close relationship with the energy spectrum of
the emergent excitations at the edge of the system in the topological phases.\cite{Li-2008,Pollmann-2010,Cirac-2011,Lou-2011,Sirker-2014}

The dynamics of the EE and the ES gives us a new perspective in the studies of the dynamical/quantum phase transitions.
For instance, it has been used to identify
the dynamical phase transitions.\cite{Vosk-2014,Canovi-2014,Torlai-2014}
Recently connection between the time evolution from the ground states and topological properties has been studied intensively.
\cite{Bermudez-2009,Bermudez-2010,Kells-2014,Hegde-2014,Mazza-2014,Nieuwenburg-2014}
In particular, dynamics during a parameter sweep across a critical point of topological phase transitions
has been investigated in the studies.
\cite{Bermudez-2010,Kells-2014,Hegde-2014}
Dynamics associated with topological phase transitions depends on their topological properties,
as pointed out in Refs.~\onlinecite{Bermudez-2009,Bermudez-2010,Kells-2014,Hegde-2014}.
In order to further understand how the topology affects the dynamics,
it is desired to know detailed properties of excited states.
It has been already recognized that
the string correlation functions, the EE, and the ES are useful to study the topological phases of the ground states.
Thus, it is an important issue to study dynamical properties in topological systems
in terms of the string correlation functions, the EE, and the ES for excited states.

In our previous letter,\cite{Ohta-2015} we have introduced a generalized cluster model 
and studied its topological phase transitions and topological properties in the sweep dynamics.
Specifically, we mapped out the ground-state phase diagram of our model,
which includes several topological and trivial phases,
and showed that there exist not only phase transitions between topological and trivial phases
but also ones between two distinct topological phases.
In addition, we briefly reported our analyses on the dynamics during interaction sweeps across the critical point
separating two topological phases with four-fold degeneracy in the ground states.
We found the breakdown of adiabaticity in this sweep dynamics even for slow sweep speed.

In this paper, we present more extensive analyses on the ground-state 
and dynamical properties of the generalized cluster model.
We show the phase diagram and detailed methods to characterize the phases.
We determine the universality class of the critical points by using the exact solution.
In addition to the case studied in Ref.~\onlinecite{Ohta-2015},
we investigate the sweep dynamics across another critical point which separates two topological phases.
To see topological properties, we calculate the ES as well as the correlation functions and the EE.
We observe spatially periodic structures in the distance dependence of correlation functions and the block-size dependence of the EE
after passing the critical points even for a slow sweep speed.
We also observe temporally oscillating and splitting structures in the time dependence of the ES
after passing the critical points.
To clarify why these structures appear,
we study the topological properties of the Bogoliubov quasiparticles (bogolons) from a viewpoint of the correlation of Majorana fermions.
In addition, we discuss the origin of the breakdown of adiabaticity.

The rest of the paper is organized as follows: 
In Sec.~\ref{sec:Preliminaries}, we introduce the model that we consider throughout this paper and 
explain the methods to analyze the model in the following sections.   
In Sec.~\ref{sec:GroundState}, we determine the ground-state phase diagram and the nature of the quantum phase 
transitions among the phases with a combination of various analytical and numerical methods.  
Paying particular attention to the bulk--edge correspondence,  
we also characterize each phase separated by the critical points with the winding number calculated from 
the bulk Hamiltonian and the degeneracy structure of the ES.
In Sec.~\ref{sec:Dynamics}, we investigate the dynamics during interaction sweeps across the critical points.
Characteristic structures in the dynamics are observed,
which can be understood from a viewpoint of the bogolons.
In Sec.~\ref{sec:Conclusion}, we conclude and summarize this paper.
In Appendices \ref{sec:operator-content} and \ref{sec:dimermodel},
we give a supplementary explanation of the results shown in Sec.~\ref{sec:UniversalityClass} and Sec.~\ref{sec:oscillating}, respectively. 

\section{\label{sec:Preliminaries}Preliminaries}
\subsection{\label{sec:Model}Model}
Throughout this paper, we consider the one-dimensional generalized cluster model introduced 
in Ref.~\onlinecite{Ohta-2015}
to study the criticality of quantum phase transitions and the dynamical properties of systems in which topological phase transitions occur.
The model includes the three-spin interaction $\sigma_{i}^x\sigma_{i+1}^z\sigma_{i+2}^x$,
the Ising interaction $\sigma_{i}^y\sigma_{i+1}^y$, and
another three-spin interaction $\sigma_{i}^y\sigma_{i+1}^z\sigma_{i+2}^y$.
The Hamiltonian is defined by
\begin{equation}
\label{eq:HGC}
H_{\rm GC}=\sum_{i=1}^{N}(-J^{XZX}\sigma_i^x\sigma_{i+1}^z\sigma_{i+2}^x+J^{YY}\sigma_i^y\sigma_{i+1}^y+J^{YZY}\sigma_i^y\sigma_{i+1}^z\sigma_{i+2}^y),
\end{equation}
where $N$ is the system size and $\sigma_i^{\alpha_0}$ $(\alpha_0=x,y,z)$ are the Pauli matrices at site $i$. 
The open boundary condition corresponds to taking $\sigma_{N+1}^{\alpha_0}=\sigma_{N+2}^{\alpha_0}=0$ ($\alpha_0 = x,y,z$)
whereas the periodic boundary condition to $\sigma_{N+1}^{\alpha_0} = \sigma_1^{\alpha_0}$ and $\sigma_{N+2}^{\alpha_0} = \sigma_2^{\alpha_0}$.   
When $J^{YZY}=0$, the model reduces to the cluster-Ising model studied in Refs.~\onlinecite{Son-2011,Smacchia-2011}.
We can flip the sign of $J^{YY}$ by the following unitary transformation:
\begin{equation}
\begin{split}
& U^{\dagger} \sigma^{x}_{i}U \mapsto (-1)^{i} \sigma^{x}_{i} \; , \;\; 
U^{\dagger}\sigma^{y}_{i} U \mapsto (-1)^{i} \sigma^{y}_{i} , \\
& U= \exp\left( {\mathrm i} \frac{\pi}{2} \sum_{j}\sigma^{z}_{j} \right) \;  .
\end{split}
\label{eqn:Jyy-flip}
\end{equation}
In general, the model with only one of the three couplings $(J^{XZX},J^{YY},J^{YZY})$ being non-zero 
possesses a special property; all the terms of the Hamiltonian are commuting and the entire spectrum is constructed 
by creating local (non-dispersive) excitations one by one.  

To clarify the physical picture of the ground state,
we explain each term of the model (\ref{eq:HGC}).
The first term in Eq.~(\ref{eq:HGC}) is called the cluster interaction or the cluster stabilizer in quantum-information science.
In the ground state of the model with $J^{YY}=J^{YZY}=0$ and positive $J^{XZX}$,
the string order parameter $O_{XZX}=\lim_{L \to \infty} O_{XZX}(L)$ is unity,
where
\begin{equation}
\label{eq:stringcorrelation}
O_{XZX}(L)=(-1)^L \left\langle\sigma_1^x\sigma_2^y\left(\prod_{i=3}^{L-2}\sigma_i^z\right)\sigma_{L-1}^y\sigma_L^x\right\rangle
\end{equation}
is called the string correlation function of distance $L$.\cite{Nijs-1989,Kennedy-1992,Smacchia-2011}
The phase characterized by the non-vanishing string order parameter is generally called the cluster (C) phase.
In the ground state of the model with $J^{XZX}=J^{YZY}=0$ and positive $J^{YY}$,
the antiferromagnetic (AF) order parameter $O_{YY}=\lim_{L \to \infty} O_{YY}(L)$ is unity,
where
\begin{equation}
\label{eq:spincorrelation}
O_{YY}(L)=(-1)^{L-1}\left\langle\sigma_1^y\sigma_L^y\right\rangle
\end{equation}
is called the spin correlation function of distance $L$.
The phase characterized by the non-vanishing AF order parameter is the AF phase. 

The last term ($J^{YZY}$) in Eq.~(\ref{eq:HGC}) is similar to the cluster interaction.
With the term there appears another topological phase, which we call the dual cluster (C*) phase.
The phase is characterized by the dual string order parameter $O_{YZY}=\lim_{L \to \infty} O_{YZY}(L)$,
where
\begin{equation}
\label{eq:dualstringcorrelation}
O_{YZY}(L)=\left\langle\sigma_1^y\sigma_2^x\left(\prod_{i=3}^{L-2}\sigma_i^z\right)\sigma_{L-1}^x\sigma_L^y\right\rangle
\end{equation}
is called the dual string correlation function of distance $L$ and characterizes the C* phase 
(see Sec.~\ref{sec:CriticalPoints}).

\subsection{\label{sec:ExactDiagonalization}Diagonalizing the Hamiltonian}
We solve the model (\ref{eq:HGC}) with the exact diagonalization method.\cite{Lieb-1961}
The original spin model (\ref{eq:HGC}) is transformed into a quadratic Hamiltonian in the spinless fermion operators \{$c_i$\}
\begin{equation}
\label{eq:Hquad}
H=\sum_{i, j=1}^N\left[c_i^{\dagger}A_{ij}c_j+\frac{1}{2}\left(c_i^{\dagger}B_{ij}c_j^{\dagger} +c_iB_{ji}c_j   \right) \right]
\end{equation}
by the Jordan--Wigner transformation:
\begin{equation}
c_i=\prod_{j=1}^{i-1}(-\sigma_j^z)\,\sigma_i^-,\quad c_i^{\dagger}=\prod_{j=1}^{i-1}(-\sigma_j^z)\,\sigma_i^+,  
\label{eqn:JW-tr}
\end{equation}
where $A$ is a real symmetric matrix of order $N$ and $B$ is a real antisymmetric matrix of order $N$.
In general, the Hamiltonian (\ref{eq:Hquad}) can be diagonalized as
\begin{equation}
\label{eq:HGCBogoliubov}
H=\sum_{\alpha=1}^{N} E_{\alpha}\left(\eta_{\alpha}^{\dagger}\eta_{\alpha}-\frac{1}{2}\right),\quad E_{\alpha}\geq0
\end{equation}
by the Bogoliubov transformation:
\begin{equation}
\label{eq:BogoliubovOperator}
\eta_{\alpha}=\sum_{i=1}^{N}\left[\frac{\phi_{i\alpha}+\psi_{i\alpha}}{2}c_i+\frac{\phi_{i\alpha}-\psi_{i\alpha}}{2}c_i^{\dagger}  \right].
\end{equation}
The matrices $\phi$ and $\psi$ of order $N$ are solution of simultaneous equations:
\begin{subnumcases}
{}
E_{\alpha}\psi_{i\alpha}=\sum_{j=1}^{N}(A+B)_{ij}\phi_{j\alpha}, \\
E_{\alpha}\phi_{i\alpha}=\sum_{j=1}^{N}(A-B)_{ij}\psi_{j\alpha}.
\end{subnumcases}
Here, the eigenenergies in Eq.~(\ref{eq:HGCBogoliubov}) are labeled in ascending order; $E_1 \le E_2 \le \cdots \le E_N$.
The Bogoliubov vacuum $\left| {\rm vac} \right\rangle$ satisfying
\begin{equation}
\eta_{\alpha} \left| {\rm vac} \right\rangle = 0, \quad \alpha=1, \dots, N
\end{equation}
is a ground state.
In our model (\ref{eq:HGC}), the matrix elements of $A$ and $B$ are given by
$A_{i,i+1}=A_{i+1,i}=J^{YY}$, $A_{i,i+2}=A_{i+2,i}=J^{XZX}-J^{YZY}$,
$B_{i,i+1}=-B_{i+1,i}=-J^{YY}$, $B_{i,i+2}=-B_{i+2,i}=J^{XZX}+J^{YZY}$, and $0$ otherwise.

\subsection{\label{sec:MajoranaFermion}Majorana fermions}
\begin{figure}
\begin{center}
  \includegraphics[scale=1]{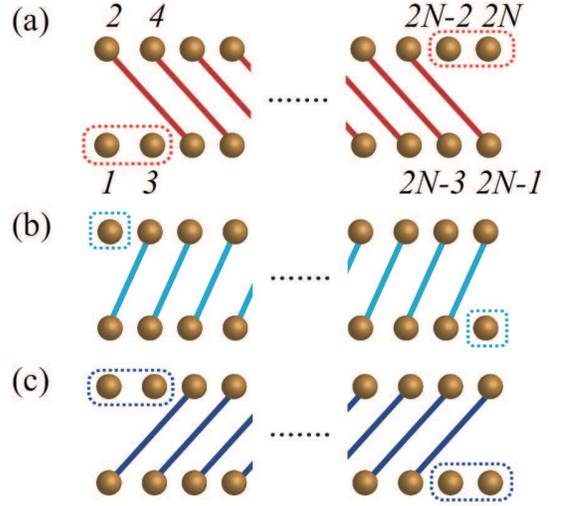} \\
\end{center}
\caption{
(Color Online)
(a)-(c) Schematic representation of the interactions in Eq.~(\ref{eq:HGC}) by the Majorana language.
(a), (b), and (c) depict the first ($J^{XZX}$), second ($J^{YY}$), and third ($J^{YZY}$) terms 
of the Hamiltonian (\ref{eq:HGC}).
Non-interacting Majorana fermions enclosed in the dotted lines appear at the ends of the system.
}
\label{fig:Fig1}
\end{figure}

To see the topological nature of the model (\ref{eq:HGC}),
we introduce the Majorana fermions.\cite{Kitaev-2001}
The Majorana fermions \{$\bar{c}_i$\} are defined by real and imaginary parts of the two fermion operators \{$c_i$\} and \{$c_i^{\dagger}$\};
a fermion is decomposed into two Majorana fermions at each site as
\begin{equation}
\bar{c}_{2i-1}=c_{i}^{\dagger}+c_i,\quad \bar{c}_{2i}=\mathrm{i}\,(c_i-c_{i}^{\dagger}),\quad i=1, 2, \dots, N.
\end{equation}
The standard anticommutation relations of \{$c_i$\} and \{$c_i^{\dagger}$\} translate into
\begin{equation}
\bar{c}_{i}=\bar{c}_{i}^{\dagger},\quad \{\bar{c}_{i},\ \bar{c}_{j}\}=2\delta_{ij}.
\end{equation}
With the basis of the Majorana fermions $\bar{c}=(\bar{c}_{1},\bar{c}_{2},\dots,\bar{c}_{2N})^{\mathrm{T}}$,
the model (\ref{eq:HGC}) is rewritten as
\begin{equation}
\label{eq:HGCmajorana}
H_{\rm GC}=\frac{\mathrm{i}}{2}\bar{c}^{\mathrm{T}}M\bar{c},
\end{equation}
where $M$ is a real antisymmetric matrix of order $2N$ with the matrix elements
$M_{2i-1,2i+2}=-M_{2i+2,2i-1}=-J^{YY}, M_{2i-1,2i+4}=-M_{2i+4,2i-1}=-J^{YZY}, M_{2i,2i+3}=-M_{2i+3,2i}=J^{XZX}$,
and $0$ otherwise.
In this representation, the coefficients of the Bogoliubov transformation $\phi$ and $\psi$ in Eq.~(\ref{eq:BogoliubovOperator})
are the amplitudes of the Majorana fermions.
With the open boundary condition,
the ground state possesses Majorana zero modes at the ends of the system.
In Figs.~\ref{fig:Fig1}(a), (b), (c),
we illustrate the interactions of the Majorana fermions represented with the yellow circles.
The colored bonds connecting them mean the interactions between the Majorana fermions;
they are paired with.
At the ends of the system,
we can see unpaired Majorana fermions enclosed by the dotted lines.
They form zero-energy fermions.

Let us consider the cases where one of the coupling constants in Eq.~(\ref{eq:HGC}) is dominant.
When either $J^{XZX}$ or $J^{YZY}$ is dominant,
we have two unpaired Majorana fermions at each end as shown in Figs.~\ref{fig:Fig1}(a), (c).
They form two zero-energy modes localized at each end of the system.
As a result, the ground states are four-fold degenerate.
On the other hand, when $J^{YY}$ is dominant,
we have two unpaired Majorana fermions as shown in Fig.~\ref{fig:Fig1}(b).
They form one zero mode,
which contributes to the two-fold degeneracy in the ground states.
The phases with zero modes at the ends of the system are called the topological phases.
The number of zero modes characterizes each phase from a topological viewpoint.\cite{Kitaev-2001,Fidkovski-2011}

\begin{figure}
\begin{center}
  \includegraphics[scale=1]{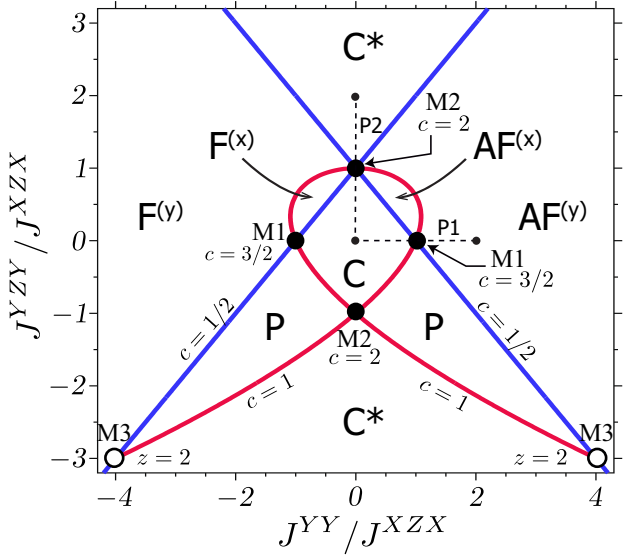} \\
\end{center}
\caption{
(Color Online)
Ground-state phase diagram of the generalized cluster model (\ref{eq:HGC}) for $J^{XZX}>0$.\cite{Ohta-2015}
On the thick solid curves, the excitation energy $E_k$ vanishes at certain values of $k$.  
The abbreviations mean the cluster (C), dual cluster (C*), ferromagnetic (F), and antiferromagnetic (AF) 
phases.
The quantum paramagnetic (P) phase cannot be characterized by string and (anti) ferromagnetic order parameters.
The superscript represents the direction of the order. 
On the blue (red) phase boundaries, a second-order transition with $c=1/2$ ($c=1$) occurs.
M1 and M2 denote multicritical points characterized by higher central charges.
The two points M3 are non-Lorentz-invariant critical points.
\label{fig:Fig2}}
\end{figure}

\subsection{\label{sec:Methods}Methods}
\subsubsection{\label{sec:CorrelationFunctions}Correlation functions}
The topological phases are characterized by non-local correlation functions.\cite{Nijs-1989,Kennedy-1992,Smacchia-2011}
Here we show how to calculate the string correlation function $O_{XZX}(L)$ as an example.
The other correlation functions are obtained in the same way.
For later convenience, we define the operators ${\cal A}$ and ${\cal B}$ as
\begin{equation}
\begin{split}
{\cal A}_i=c_i+c_i^{\dagger}=\bar{c}_{2i-1}, \\
{\cal B}_i=c_i-c_i^{\dagger}=-{\mathrm i}\, \bar{c}_{2i}.
\end{split}
\label{eqn:def-A-B}
\end{equation}
Using these operators, the string correlation function of distance $L$ is rewritten as
\begin{equation}
\label{eq:stringAB}
O_{XZX}(L)=(-1)^{L+1}\left\langle {\cal B}_1{\cal B}_2\sum_{j=3}^{L-2}\left({\cal A}_{j}{\cal B}_{j}\right){\cal A}_{L-1}{\cal A}_{L}\right\rangle,  
\end{equation}
where $\left\langle \cdot \right\rangle$ denotes the expectation value taken with respect to the Bogoliubov vacuum 
$|\text{vac}\rangle$.
This expression consists of ($2L-4$) fermion operators.
To calculate this expectation value, we need the following contractions
\begin{align}
\langle {\cal A}_i{\cal A}_j\rangle&=\delta_{ij}, \\
\langle {\cal B}_i{\cal B}_j\rangle&=-\delta_{ij}, \\
\langle {\cal B}_i{\cal A}_j\rangle&=\sum_{\alpha=1}^N\psi_{i\alpha}\phi_{j\alpha} =: D(i, j).
\end{align}
We need to calculate the Pfaffian of the antisymmetric matrix whose components are above contractions.\cite{Barouch-1971,Amico-2003}
Because operators ${\cal A}$ $({\cal B})$ appearing in Eq.~(\ref{eq:stringAB}) do not have the same index,
the Pfaffian is reduced to the determinant of the matrix of order ($L-2$) given as
\begin{equation}
  \left(
    \begin{array}{cccc}
      D(1, L) & D(1, 3) & \ldots & D(1, L-1) \\
      D(2, L) & D(2, 3) & \ldots &  \vdots \\
      \vdots & \vdots & \ddots & \vdots \\
      D(L-2, L) & \ldots & \ldots & D(L-2, L-1)
    \end{array}
  \right).
\end{equation}
%

\subsubsection{\label{sec:TimeBogoliubov}Time-dependent Bogoliubov theory}
We show the method to study the time evolution of the system.\cite{Barouch-1970,Caneva-2007}
In the following, we consider a time-dependent Hamiltonian
which is quadratic in the fermions \{$c_i$\}:
\begin{equation}
H(t)=\sum_{i, j=1}^N\left[c_i^{\dagger}A_{ij}(t)c_j+\frac{1}{2}\left(c_i^{\dagger}B_{ij}(t)c_j^{\dagger} +c_iB_{ji}^*(t)c_j   \right) \right].
\end{equation}
Here, the coupling constants depend on time $t$.
Time evolution is performed in the Heisenberg representation.
The fermions \{$c_{i,{\rm H}}(t)$\} in the Heisenberg representation obey the Heisenberg equations of motion given by
\begin{equation}
\label{eq:HEOM}
{\mathrm i}\frac{d}{dt}c_{i, \rm H}(t)=\sum_{j=1}^N\left(A_{ij}(t)c_{j, \rm H}(t)+B_{ij}(t)c_{j, \rm H}^{\dagger}(t)   \right).
\end{equation}
Let us define the Bogoliubov operators $\{\eta_{\alpha}^{\rm in}\}$
which diagonalize the Hamiltonian $H(t^{\rm in})$ at the initial time $t^{\rm in}$,
and the corresponding eigenvectors are represented as $\phi_{\alpha}^{\rm in}$ and $\psi_{\alpha}^{\rm in}$.
With these eigenvectors we define the vectors
\begin{equation}
u_{\alpha}^{\rm in}=\frac{\phi_{\alpha}^{\rm in}+\psi_{\alpha}^{\rm in}}{2},\quad v_{\alpha}^{\rm in}=\frac{\phi_{\alpha}^{\rm in}-\psi_{\alpha}^{\rm in}}{2}
\end{equation}
for later convenience.
The Heisenberg equation of motion (\ref{eq:HEOM}) is solved in the following way:
Let us write fermions $c_{i, \rm H}(t)$ as
\begin{equation}
c_{i, \rm H}(t)=\sum_{\alpha=1}^N\left(u_{i, \alpha}(t)\eta_{\alpha}^{\rm in}+v_{i, \alpha}^*(t)\eta_{\alpha}^{\rm in\dagger}   \right),
\end{equation}
where $u_{i,\alpha}(t)$ and $v_{i,\alpha}(t)$ denote the $i$-th component of $u_\alpha(t)$ and $v_\alpha(t)$, respectively.
By substituting this expression for Eq.~(\ref{eq:HEOM})
we obtain the simultaneous linear differential equations of $u_{\alpha}(t)$ and $v_{\alpha}(t)$
\begin{subnumcases}
{\label{eq:SLDEuv}}
{\mathrm i}\frac{d}{dt}u_{i, \alpha}(t)=\sum_{j=1}^N\left(A_{ij}(t)u_{j, \alpha}(t)+B_{ij}(t)v_{j, \alpha}(t)   \right), \\
{\mathrm i}\frac{d}{dt}v_{i, \alpha}(t)=-\sum_{j=1}^N\left(A_{ij}(t)v_{j, \alpha}(t)+B_{ij}(t)u_{j, \alpha}(t)   \right)
\end{subnumcases}
with the initial condition
\begin{equation}
u_{\alpha}(t_{\rm in})=u_{\alpha}^{\rm in},\quad v_{\alpha}(t_{\rm in})=v_{\alpha}^{\rm in}.
\end{equation}
With the above setup, we calculate the expectation value of an operator $O(c_i, c_i^{\dagger})$ at time $t$
\begin{equation}
\langle\Psi(t)|O(c_i, c_i^{\dagger})|\Psi(t)\rangle=\langle\Psi(t_{\rm in})|O(c_{i, \rm H}(t), c_{i, \rm H}^{\dagger}(t))|\Psi(t_{\rm in})\rangle.
\end{equation}
When the initial state $|\Psi(t_{\rm in})\rangle$ is set to the Bogoliubov vacuum,
the expectation value can be calculated as in the case of the time-independent Hamiltonian.
Because only $u_{\alpha}(t)$ and $v_{\alpha}(t)$ depend on time $t$,
what we need to do is to calculate the time evolution of $u_{\alpha}(t)$ and $v_{\alpha}(t)$ in Eqs.~(\ref{eq:SLDEuv}).

\section{\label{sec:GroundState}Ground-state properties}
\subsection{\label{sec:CriticalPoints}Phases and critical points}
To determine the phase boundaries of the ground-state phase diagram of the model (\ref{eq:HGC}),
we calculate the energy spectrum under
the periodic boundary condition.
By performing the Fourier transformation followed by the Bogoliubov transformation, the model is expressed as
\begin{equation}
\label{eq:BdG}
H=\sum_{0 \leq k \leq \pi}E_k\,(\eta_k^{\dagger}\eta_k+\eta_{-k}^{\dagger}\eta_{-k}),\quad
E_k = 2\sqrt{\epsilon_k^2+\delta_k^2},
\end{equation}
where $E_k\geq0$ is the excitation energy at the wave number $k$ and
\begin{align}
\epsilon_k&=(J^{XZX}-J^{YZY})\cos 2k+J^{YY}\cos k, \\
\delta_k&=(J^{XZX}+J^{YZY})\sin 2k-J^{YY}\sin k.
\end{align}
In the following, we use $J^{XZX}$ as the energy unit and
only consider the case with positive $J^{XZX}$ for concreteness.

When the periodic boundary condition is imposed, the unpaired Majorana modes disappear 
and the degeneracy in the topological phase occurs due to the existence of the zero modes.  
The Hamiltonian in the momentum space has the form of the Bogoliubov--de Gennes Hamiltonian with time reversal symmetry 
and the ground state is given by the vacuum of the Bogoliubov operators $\{\eta_k\}$. 
The critical points which separate phases are determined by the condition that the excitation energy $E_k$ vanishes 
at certain wave numbers $k$.
They are shown by the thick solid lines in Fig.~\ref{fig:Fig2}. 

Once the phase boundaries are determined, we can identify the phases by 
calculating the order parameters with the time-evolving block decimation method for infinite systems (iTEBD)
or by considering the extreme cases and using the adiabatic continuity.\cite{Ohta-2015}
For instance, the phase ``P'' is continuously connected to the quantum paramagnetic phase where 
$\sigma^{z}$ are polarized. 
As shown in Sec.~\ref{sec:Model},
the order parameters $O_{XZX}$, $O_{YZY}$, and $O_{YY}$ have finite values in the C, C*, and AF$^{(y)}$ phases, respectively.
The antiferromagnetic phase in the $x$ direction (AF$^{(x)}$) is characterized by the AF order parameter $O_{XX}$
\begin{equation}
\label{eq:SpinOrderxx}
O_{XX}=\lim_{L \to \infty}(-1)^{L-1}\left\langle\sigma_1^x\sigma_L^x\right\rangle.
\end{equation}
The results are summarized in Fig.~\ref{fig:Fig2}.
The phases on the right of the $J^{YZY}$-axis and the ones on the left are mapped onto each other 
by the unitary transformation \eqref{eqn:Jyy-flip};
the AF phases are mapped onto the ferromagnetic (F) phases.
The AF$^{(x)}$, F$^{(x)}$, and P phases appear as a result of competing interactions in the model (\ref{eq:HGC}).
We can see that all the phase transitions are continuous from the order parameters (not shown here).

\subsection{\label{sec:WindingNumber}Winding number}
The ground state of the Hamiltonian (\ref{eq:BdG}) is characterized by a topological invariant called the winding number.\cite{Anderson-1958,Niu-2012}
The Hamiltonian is rewritten as
\begin{equation}
  H=\sum_{k} (c_k^{\dagger},\ c_{-k}) H(k) (c_k,\ c_{-k}^{\dagger})^{\mathrm{T}},\quad H(k) 
  = \mbox{\boldmath $d$}(k)\cdot \boldsymbol{\sigma}
\end{equation}
by using the Anderson pseudospin $\mbox{\boldmath $d$}(k)$
\begin{equation}
\mbox{\boldmath $d$}(k)=\epsilon_k\hat{\mbox{\boldmath $e$}}_z+\delta_k\hat{\mbox{\boldmath $e$}}_y,
\end{equation}
where $\hat{\mbox{\boldmath $e$}}_y$, $\hat{\mbox{\boldmath $e$}}_z$ are the unit vectors in the $y$, $z$ directions, respectively,
and $\boldsymbol{\sigma}=(\sigma^{x},\sigma^{y},\sigma^{z})$ represents the Pauli matrices.
We define the angle in the $yz$-plane by using the normalized Anderson pseudospin:
\begin{equation}
\hat{\mbox{\boldmath $d$}}(k)=\frac{\mbox{\boldmath $d$}(k)}{|\mbox{\boldmath $d$}(k)|}=\cos\theta_k\hat{\mbox{\boldmath $e$}}_z+\sin\theta_k\hat{\mbox{\boldmath $e$}}_y.
\end{equation}
The angle $\theta_k$ maps one-dimensional Brillouin Zone (circle) to the Hilbert space (circle).
Then we can define the winding number $W$ as the topological invariant of the mapping
\begin{equation}
W=\int_{\rm B.Z.}\frac{d\theta_k}{2\pi} =
\begin{cases}
+2 &({\rm C \,\,phase})\\
-2 & ({\rm C^* \,\,phase})\\
+1 & ({\rm F}^{(x)}/{\rm AF}^{(x)} \,\,{\rm phases})\\
-1 & ({\rm F}^{(y)}/{\rm AF}^{(y)} \,\,{\rm phases})\\
0 & ({\rm P \,\,phase})
\end{cases}.
\end{equation}
The winding number has an integer value and changes only if the system becomes critical.
Therefore each phase has definite value of the winding number.
These numbers correspond to the number of zero modes appearing at the edge of the system when the open boundary condition is imposed.

\subsection{\label{sec:EntanglementProperties}Entanglement properties}
There is yet another way to characterize the phases found in Sec.~\ref{sec:CriticalPoints} using 
quantum entanglement.  
To this end, we define two quantities, the EE and the ES, which quantify entanglement.\cite{Li-2008,Latorre-2004}
We divide the entire system into a subsystem A
with the length $L_{\rm sub}$ and the rest B in a way symmetric with respect to the center of the system 
and measure the entanglement between them.   
We calculate the eigenvalues \{$\lambda_{\beta}$\} of the reduced density matrix $\rho_{\rm A}$ of A  
which is obtained from the density matrix of the entire system $\rho$ by tracing out the subsystem B:
\begin{equation}
  \label{eq:rdm}
\rho_{\rm A}={\rm Tr}_{\rm B} \, \rho \; .
\end{equation}
With the method developed in Refs.~\onlinecite{Peschel-2003,Peschel-2009},
one obtains the reduced density matrix from the fermionic correlation functions for the excited states as well as the ground states.
We define the EE as the von Neumann entropy of the reduced density matrix $\rho_{\rm A}$:
\begin{equation}
  \label{eq:ee}
S=-{\rm Tr} (\rho_{\rm A}\ln \rho_{\rm A}).
\end{equation}
On the other hand, the ES is defined as
\begin{equation}
  \label{eq:es}
\xi_{\beta} = -\ln \lambda_{\beta},
\end{equation}
where $\{\lambda_\beta\}$ are the eigenvalues of $\rho_{\rm A}$.\cite{Li-2008}
Although the EE and the ES are well-defined in the analysis of only the ground state,
we expect that the EE and the ES defined for the excited states can extract some physical properties in the excited states.

We characterize the phases in Fig.~\ref{fig:Fig2} by the ES.\cite{Ohta-2015}
Here the open boundary condition is imposed.
We examine the number of degeneracy of the lowest level in the ES.
The number of degeneracy is the same as that of the ground states originating from the Majorana zero modes 
at the ends of the system.  
We thus confirm that for the generalized cluster model (\ref{eq:HGC})
the ES reflects the fictitious degree of freedom appearing at the cut ends
as in the case of other topological phases studied in previous works.
\cite{Li-2008,Pollmann-2010,Sirker-2014}

Table~\ref{table:BE} summarizes the relation between the winding number and the number of degeneracy of the lowest level in the ES. 
The former is calculated with the periodic boundary condition, while 
the latter is calculated with the open boundary condition and reflects the edge modes.
We can clearly see that
the absolute value of the winding number is equal to the number of zero-energy excitations 
in an open chain.  
This is a manifestation of the bulk--edge correspondence in the model \eqref{eq:HGC}.
Note that the winding number or the degree of degeneracy of the ES alone does not tell anything about physical properties 
of the phases except for topologically defined numbers (e.g., the number of edge excitations).
To determine the phases completely, we need to calculate the order parameters by other methods 
as we did in Sec.~\ref{sec:CriticalPoints}.

\begin{table}
  \caption{The relation between the winding number and the number of degeneracy of the ES 
  (bulk--edge correspondence). Dominant order parameter (OP) for each phase is also shown. }
  \label{table:BE}
\begin{ruledtabular}
  \begin{tabular}{cccc}
    Phase  & OP & Winding number  &  Degeneracy of the ES  \\
    \hline 
    C  & $O_{XZX}$ & $2$  & four-fold \\
    C*  & $O_{YZY}$ & $-2$   & four-fold \\
    ${\rm F}^{(x)}/{\rm AF}^{(x)}$ & $O_{XX}$ & $1$  & two-fold \\
    ${\rm F}^{(y)}/{\rm AF}^{(y)}$  & $O_{YY}$ & $-1$  & two-fold \\
    P  & -- & $0$  &  no degeneracy \\
  \end{tabular}
  \end{ruledtabular}
\end{table}

\subsection{\label{sec:UniversalityClass}Universality class}
For completeness, let us determine the universality classes of the transitions. 
In fact, all the phase transitions that occur are continuous and described by Lorentz-invariant 
conformal field theories (CFT) except at the points marked in Fig.~\ref{fig:Fig2} 
as ``M3'', where the dispersion $E_{k}$ is quadratic in $k$ and the dynamical critical exponent takes $z=2$.  
The information on the universality class of quantum phase transition 
and the corresponding central charge $c$ at each critical point can be most conveniently extracted 
from the finite-size energy spectrum\cite{Blote-1986,Cardy-1986} 
(or equivalently, from the low-temperature behavior of the free energy density\cite{Affleck-1986})
or from the scaling behavior of the block entanglement entropy.\cite{Holzhey-1994,Calabrese-2004}
In order to obtain more precise information on the universality class, we adopt the former and identify 
the contents of the scaling operators using the spectrum obtained exactly above.  
Throughout this subsection, we assume the periodic boundary condition. 

The method relies on the CFT prediction on the finite-size spectrum of a $(1+1)$-dimensional quantum 
system at the critical point\cite{Cardy-1986}:
\begin{equation}
\begin{split}
& E_{h,\bar{h}}(N) = N \epsilon_{\infty} - \frac{\pi v_{\text{s}}}{6N}c + \frac{2\pi}{N}v_{\text{s}} (h + \bar{h} 
+ n_{\text{L}} + n_{\text{R}} ) \\
&  (n_{\text{L}}, n_{\text{R}}=0,1,2, \ldots) \; ,
\end{split}
\label{eqn:CFT-FSS}
\end{equation}
where $\epsilon_{\infty}$ is the ground-state energy density in the infinite-size limit and 
$v_{\text{s}}$ is the velocity that characterizes the $k$-linear dispersion of the critical excitations. 
In general, the entire spectrum decomposes into the several sectors labeled by 
the conformal weights $(h,\bar{h})$.  
The central charge $c$ and the list of the pairs $(h,\bar{h})$ appearing 
in the actual spectrum (i.e., operator contents) determines the universality.  
As the exact spectrum is obtained in a closed form here, it is rather straightforward to obtain these data 
(see Appendix \ref{sec:operator-content} for more details).    

In Fig.~\ref{fig:Fig2}, we show the universality classes obtained in this way.  
On the blue solid lines, there is only one gapless $k$-linear Majorana point at $k=0$ or $k=\pi$.  
The quantum phase transition there belongs to the Ising universality class with $c=1/2$.  
Extracting $c$ for the phase boundaries shown by the red solid lines is tricky as the two Majorana points 
are located at incommensurate momenta and the approach to the infinite-size limit is quite irregular. 
So we used the free energy density instead and fitted the low-temperature ($T$) free energy density 
$F(T,N)/N$, which is calculated exactly using Eq.~\eqref{eq:BdG}
by its CFT asymptotic form\cite{Affleck-1986} 
\begin{equation}
\frac{1}{N} F(T,N) \sim \epsilon_{\infty} - \frac{\pi c}{6 v_{\text{s}}} T^{2} 
\end{equation}
to obtain $c=1$.   

At the multicritical points M1: $(J^{YY}/J^{XZX},J^{YZY}/J^{XZX})=(\pm1, 0)$, the system has three 
gapless $k$-linear Majorana points 
at $k=0$, $\pm2\pi/3$ (or at $k=\pm \pi/3$, $\pi$) and, by fitting the ground-state energy to the scaling form 
\eqref{eqn:CFT-FSS}, we readily obtain $c=3/2$.  However, there are several different universality 
classes with $c=3/2$ and we need more precise analysis.
Using the method sketched in Appendix \ref{sec:operator-content}, we see that 
the universality class is the level-1 $\text{SO}(3)$ Wess--Zumino--Witten (WZW) model with $c=3/2$.\cite{comment-WZW}
A similar argument shows that 
the points $(J^{YY}/J^{XZX},J^{YZY}/J^{XZX})=(0, \pm1)$ (M2) correspond to the universality class of 
the level-1 $\text{SO}(4)$ WZW with $c=2$ with four Majorana points at $k=0$, $\pm \pi/2$, and $\pi$ 
(or at $k=\pm \pi/4, \pm 3\pi/4$).
These two SO($N$) criticalities fit into the series of general SO($N$) critical points discussed in Ref. \onlinecite{Lahtinen-2015}.
At the point M3: $(J^{YY}/J^{XZX},J^{YZY}/J^{XZX})=(-4, -3)$ [$(4,-3)$] where the two second-order 
phase boundaries merge, the quasi-particle dispersion takes 
the non-relativistic form $E_{k} \sim k^{2}$ near $k=0$ [$E_{k} \sim (k-\pi)^{2}$ near $k=\pi$] 
and the continuous quantum phase transition is characterized by the dynamical exponent $z=2$ (Table~\ref{table:QPT}).

\begin{table*}
  \caption{Quantum phase transitions in the model \eqref{eq:HGC}.  For the multicritical points, the critical exponents 
  may depend on how we deviate from the critical point. In most cases, we consider the deviation 
  in the $J^{YY}$-direction. The exponent $\nu$ characterizes the spatial correlation length $\xi_{x}$ as 
  $\xi_{x} \sim |J^{YY} - J^{YY}_{\text{c}}|^{-\nu}$.}
  \label{table:QPT}
\begin{ruledtabular}
  \begin{tabular}{lccc}
    Phase boundaries & $c$ & Universality  &  Exponents  \\
    \hline 
    P/C*-F$^{(y)}$, P/C*-AF$^{(y)}$,  C-F$^{(x)}$/AF$^{(x)}$ & $1/2$ & Ising & $z=1$, $\nu=1$ \\
    C-P, C*-P, F$^{(x)}$-F$^{(y)}$, AF$^{(x)}$-AF$^{(y)}$ & $1$ &  XY & $z=1$ \\
    M1 & $3/2$  & SO(3)$_1$ WZW & $z=1$, $\nu=1$\\
    M2  & $2$  & SO(4)$_1$ WZW & $z=1$, $\nu=1$ \\
    M3  &  --  & 1D hardcore boson & $z=2$, $\nu=1$\\
  \end{tabular}
  \end{ruledtabular}
\end{table*}

\section{\label{sec:Dynamics}Sweep Dynamics}
In the previous section, the ground-state phases of the model \eqref{eq:HGC} and topological quantum 
phase transitions were investigated from the viewpoint of the entanglement and 
the bulk--edge correspondence.
In this section, we study the dynamical properties of the model (\ref{eq:HGC})
when an interaction parameter changes across these critical points at a finite speed. 
The sweep dynamics captures the effects of the low-energy excitations more clearly 
than the quench dynamics in which the parameter is changed abruptly.  
We study the dynamics under the sweep from the C phase to the AF$^{(y)}$ phase 
in Sec.~\ref{sec:CAF} and to the C* phase in Sec.~\ref{sec:CC*}.
In the former case, the ground-state degeneracy changes during the sweep.   
In the latter case, on the contrary, the ground-state degeneracy does not change in the sweep dynamics.  

In the fermion representation, all these three (AF$^{(y)}$, C, and C*) 
are the topological phases with ground-state degeneracy.  
As was pointed out in Refs.\onlinecite{Bermudez-2009,Bermudez-2010,Kells-2014,Hegde-2014},
the topological properties of the initial state strongly affect the time evolution of the system.
Since the energies of the two lowest bogolons $\eta_1$ and $\eta_2$ vanish ($E_1=E_2=0$) in the C phase,
the states $\eta_1^\dagger \left| {\rm vac} \right\rangle$, $\eta_2^\dagger \left| {\rm vac} \right\rangle$,
and $\eta_2^\dagger \eta_1^\dagger \left| \text{vac} \right\rangle$ together with 
the vacuum state $\left| {\rm vac} \right\rangle$ constitute the ground-state subspace.
In the case of the Kitaev chain,\cite{Kitaev-2001} 
the states are labeled by the fermionic parity operator defined by
\begin{equation}
\label{eq:parity}
\prod_{i=1}^{N}\sigma_i^z,
\end{equation}
which is crucial to understand the topological properties.

In the cluster model, on the other hand, the states are labeled by the following set of parity operators defined by\cite{Smacchia-2011,Bahri-2015}
\begin{equation}
\label{eq:parityC}
\prod_{i \in {\rm even}}\sigma_i^z, \quad \prod_{i \in {\rm odd}}\sigma_i^z.
\end{equation}
In fact, the above two operators are conserved anywhere along the line $J^{YY}=0$ and 
crucial in understanding the topological properties of the C phase.
In the following, the initial state is prepared in the vacuum $\left| {\rm vac} \right\rangle$
that is a ground state of the model (\ref{eq:HGC}).
Because there is an ambiguity to construct the zero modes $\eta_1$ and $\eta_2$ in the initial state,
we use the following expression for them:
\begin{equation}
\label{eq:zeromode}
\eta_1=\frac{1}{2}(\bar{c}_{1}-\mathrm{i}\bar{c}_{2N}), \quad \eta_2=\frac{1}{2}(\bar{c}_{3}-\mathrm{i}\bar{c}_{2N-2}).
\end{equation}
The eigenvalues of the fermionic parity operators (\ref{eq:parityC}) in the vacuum are both $-1$.

\subsection{\label{sec:CAF}C to AF$^{(y)}$}
\begin{figure}
\begin{center}
  \includegraphics[scale=1]{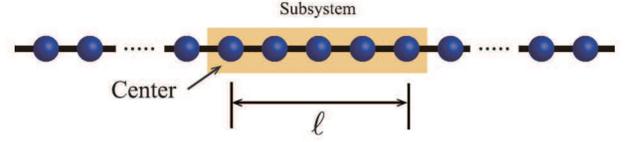} \\
\end{center}
\caption{
(Color Online)
The spin-spin and string correlation functions $\mathcal{O}_{YY}$ and $O_{YZY}$ 
are measured between the central site (site 1) and the other at a distance $\ell$.  
In calculating the EE, we take a block of $\ell$ adjacent sites to the right of the central site.
}
\label{fig:Fig3}
\end{figure}

\begin{figure}
\begin{center}
  \includegraphics[scale=1]{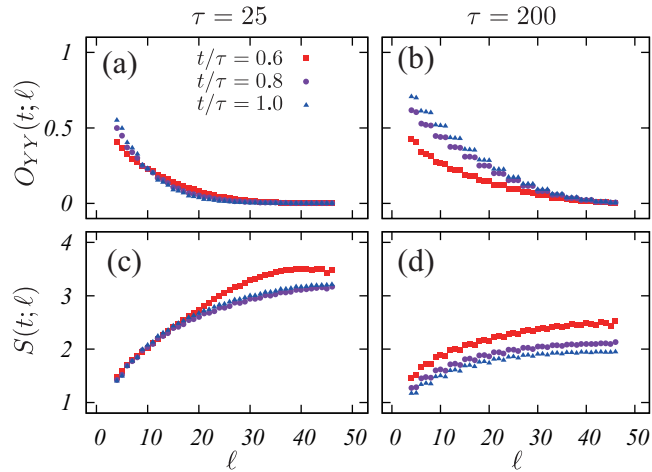} \\
\end{center}
\caption{
(Color Online)
Sweep dynamics from the C phase to the AF$^{(y)}$ phase.  
$\ell$ denotes the distance between the two end points (for the correlation function) or 
the size of the block (for the EE).  See Fig.~\ref{fig:Fig3} for more details. 
The system size is $N=101$.
(a) The distance dependence of the correlation function $O_{YY}(\ell)$ at $t/\tau=0.6, 0.8, 1$ with $\tau=$ 25.
(b) The same plot for $\tau=$ 200.
(c) The block-size dependence of the EE $S(\ell)$ at $t/\tau=0.6, 0.8, 1$ with $\tau=$ 25.
(d) The same plot for $\tau=$ 200.
}
\label{fig:Fig4}
\end{figure}

We begin by studying the dynamics during an interaction sweep across the critical point between the C and AF$^{(y)}$ phases with the open boundary condition.
In this case, the degree of degeneracy of the ground states is four and two in the C and AF$^{(y)}$ phases, respectively.
Therefore the situation is similar to the Kitaev model
where the mismatch between the degeneracies in the topological and trivial phases occurs.\cite{Kells-2014,Hegde-2014}
Let us first set $J^{YZY}=0$ and consider the following time-dependent Hamiltonian:
\begin{equation}
\label{eq:H1}
H_1(t)=-J^{XZX}\sum_{i=1}^{N} \sigma_i^x\sigma_{i+1}^z\sigma_{i+2}^x+J^{YY}(t)\sum_{i=1}^{N} \sigma_i^y\sigma_{i+1}^y,
\end{equation}
where the interaction parameter changes linearly during the sweep time $\tau$ as
\begin{equation}
J^{YY}(t)/J^{XZX}=2t/\tau,\quad 0\le t \le \tau
\end{equation}
tracing the path shown by the dashed line `P1' in Fig.~\ref{fig:Fig2}.  
At $t=0$, $H_1(t)$ coincides with the cluster model and it gradually evolves into the final form 
$(J^{YY}/J^{XZX},J^{YZY}/J^{XZX})=(2,0)$.  
Note that at $t=\tau/2$ the system passes through the multicritical point M1 to enter the AF$^{(y)}$ phase 
(see Fig.~\ref{fig:Fig2}).
We calculated the distance ($\ell$) dependence of the spin correlation function $O_{YY}(\ell)$ 
[Eq.~\eqref{eq:spincorrelation}] 
and the block-size dependence of the EE $S(\ell)$ by using the time-dependent Bogoliubov transformation explained 
in Sec.~\ref{sec:TimeBogoliubov},\cite{Barouch-1970,Caneva-2007}
where $\ell$ is defined as Fig.~\ref{fig:Fig3}. 
We show the $\ell$-dependence of the spin correlation function
at time $t/\tau=0.6, 0.8, 1$ for different sweep times $\tau=$ 25 and 200 in Figs.~\ref{fig:Fig4}(a) and (b), 
respectively.
A triple-periodic structure in the $\ell$-dependence is clearly visible for larger $\tau$, i.e., slower sweep.  
We also plot the block-site ($\ell$) dependence of the EE 
at time $t/\tau=0.6, 0.8, 1$ with $\tau=$ 25 and 200 in Figs.~\ref{fig:Fig4}(c) and (d), respectively.
We can clearly see a similar period-3 structure in the block-size dependence of the EE for the slower sweep.   

\begin{figure}
\begin{center}
  \includegraphics[scale=1]{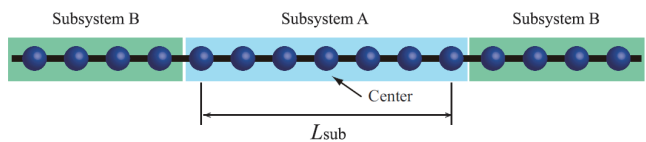} \\
\end{center}
\caption{
(Color Online)
Schematic of subsystems A and B.
}
\label{fig:Fig5}
\end{figure}

\begin{figure*}
\begin{center}
  \includegraphics[scale=1]{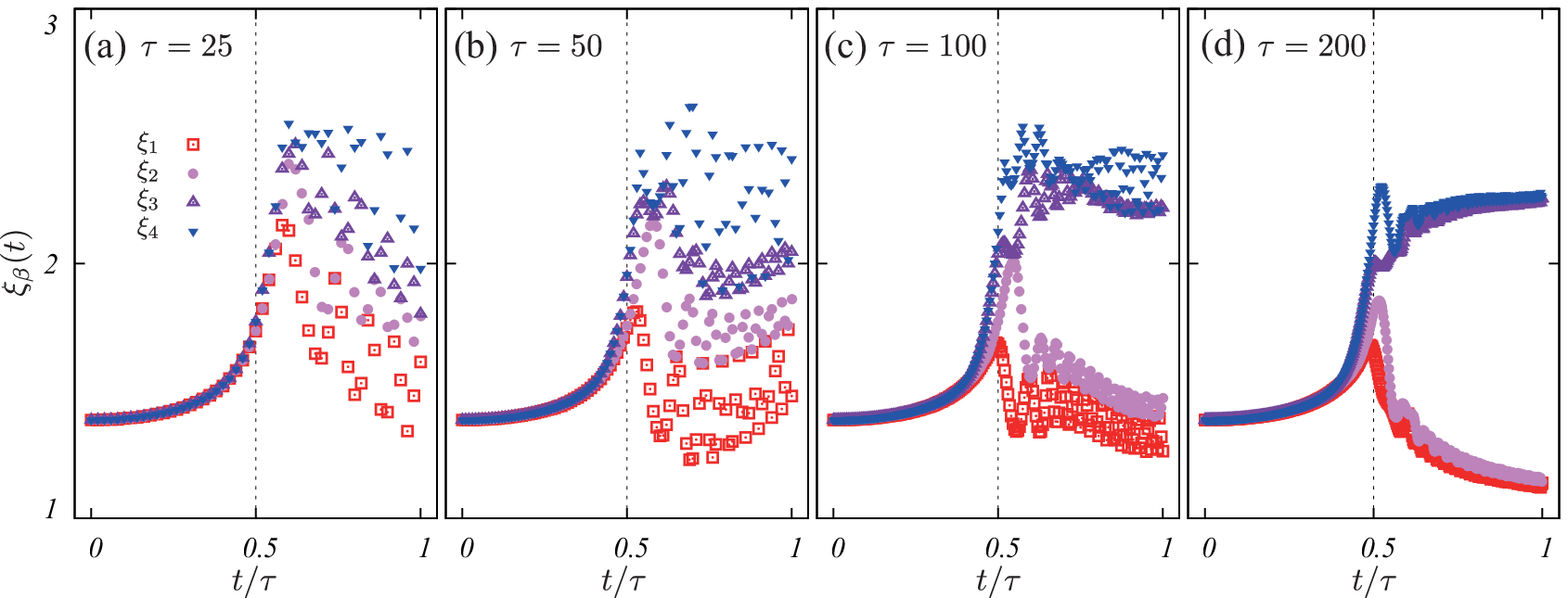} \\
\end{center}
\caption{
(Color Online)
Sweep dynamics from the C phase to the AF$^{(y)}$ phase.
We calculate the ES with $N=101$ and $L_{\rm sub}=49$ (see Fig.~\ref{fig:Fig5}). 
The time evolution of the lowest four entanglement levels $\xi_{\beta}(t)$ ($\beta=1,2,3,4$)
with (a) $\tau=25$, (b) $\tau=50$, (c) $\tau=100$, and (d) $\tau=200$.
}
\label{fig:Fig6}
\end{figure*}

To see the change in the topological properties during the sweep,
we calculated the time dependence of the ES.
We choose the subsystem A of size $L_{\rm sub}=49$ located 
symmetrically around the center of the entire system of length $N=101$ (see Fig.~\ref{fig:Fig5}).  
In Figs.~\ref{fig:Fig6}(a)-(d), we show the lowest four entanglement levels for the sweep times $\tau=25, 50, 100, 200$.
Up to the time $t=\tau/2$ when the instantaneous Hamiltonian undergoes a quantum phase transition from the C phase to the AF$^{(y)}$ phase,
the levels are four-fold degenerate.
After passing the critical point, the degeneracy resolves:
For faster sweeps the four levels oscillate in time 
[see Figs.~\ref{fig:Fig6}(a)-(c)],
whereas for slower sweeps the levels split into two pairs 
[see Fig.~\ref{fig:Fig6}(d)].

\subsection{\label{sec:CC*}C to C*}
Next we turn to the dynamics during an interaction sweep across the critical point
between the C and C* phases with the open boundary condition.
Because the degree of degeneracy of the ground states is four in both the C and C* phases,
the mismatch between the degeneracies does not occur.
Therefore, the situation differs from that in Refs.~\onlinecite{Kells-2014,Hegde-2014} 
or in Sec~\ref{sec:CAF}.
Let us now set $J^{YY}=0$ and consider the following time-dependent Hamiltonian:
\begin{equation}
\label{eq:H2}
H_2(t)=-J^{XZX}\sum_{i=1}^{N} \sigma_i^x\sigma_{i+1}^z\sigma_{i+2}^x+J^{YZY}(t)\sum_{i=1}^{N} \sigma_i^y\sigma_{i+1}^z\sigma_{i+2}^y \; .
\end{equation}
As the interaction parameter $J^{YZY}$ grows linearly during the sweep time $\tau$ as
\begin{equation}
J^{YZY}(t)/J^{XZX}=2t/\tau,\quad 0\le t \le \tau \; ,
\end{equation}
the Hamiltonian evolves from the cluster model into the one with $(J^{YY}/J^{XZX},J^{YZY}/J^{XZX})=(0,2)$. 
Note that the $c=2$ multicritical point M2 is passed at $t=\tau/2$ (see the dashed line shown as 
`P2' in Fig.~\ref{fig:Fig2}).  

\begin{figure}
\begin{center}
  \includegraphics[scale=1]{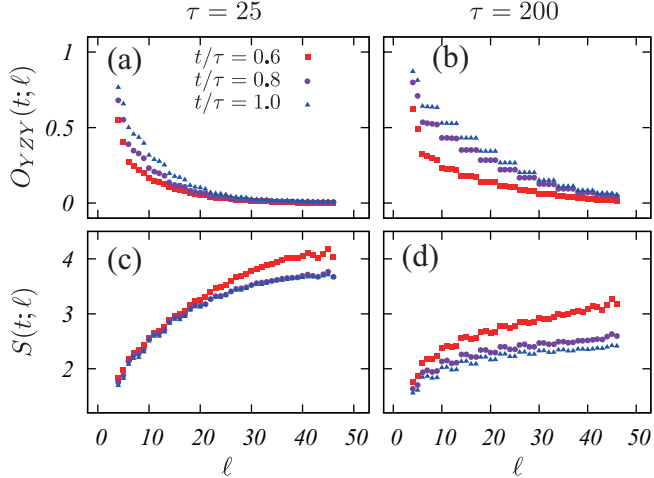} \\
\end{center}
\caption{
(Color Online)
Sweep dynamics from the C phase to the C* phase.  
$\ell$ denotes the distance between the two end points (for the correlation function) or 
the size of the block (for the EE).  
The system size is $N=101$.
(a) The distance dependence of the correlation function $O_{YY}(\ell)$ at $t/\tau=0.6, 0.8, 1$ with $\tau=25$.
(b) The same plot for $\tau=200$.
(c) The block-size dependence of the EE $S(\ell)$ at $t/\tau=0.6, 0.8, 1$ with $\tau=25$.
(d) The same plot for $\tau=200$.
}
\label{fig:Fig7}
\end{figure}

We calculated the $\ell$-dependence of the dual string correlation function $O_{YZY}(\ell)$ 
[Eq.~\eqref{eq:dualstringcorrelation}; with $\ell$ being the distance between the two end points] 
and the block EE $S(\ell)$ (with $\ell$ being the size of the block) in the same way as in the previous section.
We show the $\ell$-dependence of the dual string correlation function
at different elapsed times $t/\tau=0.6, 0.8, 1$ in Figs.~\ref{fig:Fig7}(a) and (b) 
for $\tau=$ 25 and 200, respectively.
A period-4 structure in the $\ell$-dependence is clearly seen for larger $\tau$ (slower sweep).
Next, we calculated the block-size ($\ell$) dependence of the EE 
at $t/\tau=0.6, 0.8, 1$.  The results for $\tau=$ 25 and 200 are shown in Figs.~\ref{fig:Fig7}(c) and (d), respectively. 
As in the case of the C-AF$^{(y)}$ sweep in Sec.~\ref{sec:CAF}, 
the block-size dependence of the EE exhibits a similar periodic structure to that 
of the dual string correlation function $O_{YZY}(\ell)$.   

\begin{figure*}
\begin{center}
  \includegraphics[scale=1]{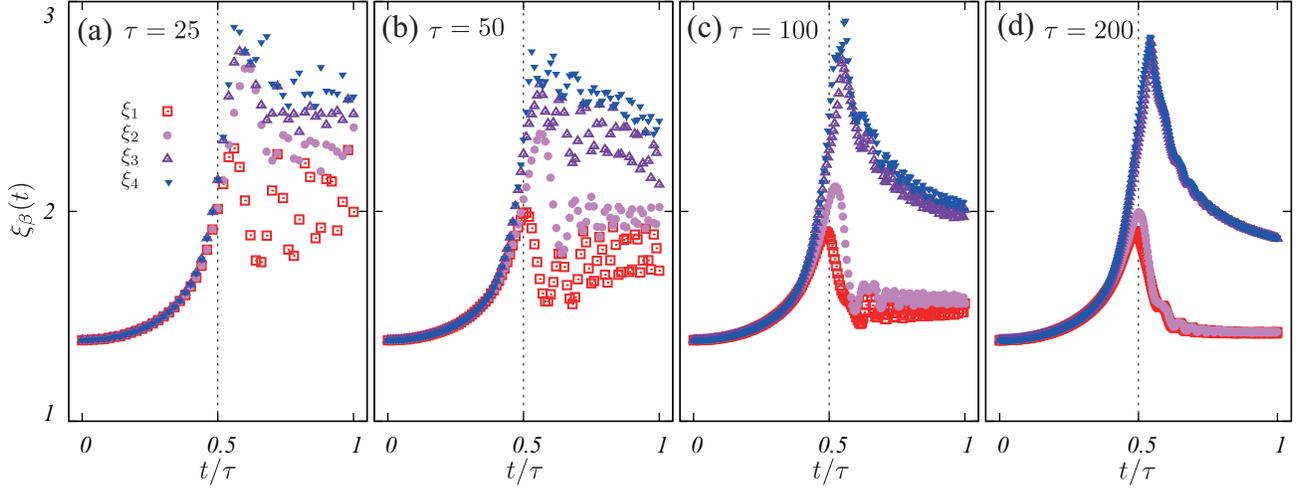} \\
\end{center}
\caption{
(Color Online)
Sweep dynamics from the C phase to the C* phase.
We calculate the ES with $N=101$ and $L_{\rm sub}=49$.
The time evolution of the lowest four entanglement levels $\xi_{\beta}(t)$ ($\beta=1,2,3,4$)
with (a) $\tau=25$, (b) $\tau=50$, (c) $\tau=100$, and (d) $\tau=200$.
}
\label{fig:Fig8}
\end{figure*}

We calculated the time-evolution of the ES.   
We take the same subsystem as in the previous section (see Fig.~\ref{fig:Fig5}). 
In Figs.~\ref{fig:Fig8}(a)-(d), we plot the lowest four entanglement levels for different sweep times 
$\tau=25, 50, 100, 200$.  
Up to the time $t=\tau/2$, when the instantaneous Hamiltonian is located at the multicritical point M2 
from the C phase to the C* phase, the levels retain the four-fold degeneracy.
After passing the critical point, the degeneracy is lifted:
For faster sweeps, the four levels oscillate in time
[see Figs.~\ref{fig:Fig8}(a)-(b)],
while for slower sweeps the levels split into pairs  
[see Fig.~\ref{fig:Fig8}(d)].

\subsection{\label{sec:periodic}Periodic structure in correlation functions and the EE}
To elucidate the origin of the periodic structure in the correlation functions and the EE 
found in the sweep dynamics shown in the previous sections,  
we calculate the correlation functions and the EE in certain excited states.
Since it is expected that for slow sweeps the main contribution to the dynamical properties 
comes from low-lying excited states,\cite{Ohta-2015} we may focus only on the states with a single bogolon.  
That is, we consider only the excited states that are generated by applying a single bogolon operator 
$\eta^{\dagger}_{\alpha}$ to the Bogoliubov vacuum:
\begin{equation}
|\alpha\rangle=\eta_{\alpha}^{\dagger}|{\rm vac}\rangle \quad (\alpha=1, 2, \dots, N), 
\end{equation}
where the Bogoliubov energy $E_{\alpha}$ is the eigenvalue of the instantaneous Hamiltonian 
at some given time $t$ and is assumed to be labeled in the ascending order.  
The expression of the string correlation function $O_{XZX}(\ell)$ calculated for the excited states 
$|\alpha\rangle$ [see Eq.~\eqref{eq:stringAB}]
\begin{equation}
O_{XZX}(\ell)_{\alpha}
\equiv 
\left\langle \eta_{\alpha} \, \left\{
{\cal B}_1{\cal B}_2\sum_{j=3}^{\ell-2}\left({\cal A}_{j}{\cal B}_{j}\right){\cal A}_{\ell-1}{\cal A}_{\ell} \right\} \, 
\eta_{\alpha}^{\dagger}
\right\rangle
\end{equation}
contains ($2\ell-2$) fermion operators ${\cal A}$ and ${\cal B}$ 
[see Eq.~\eqref{eqn:def-A-B} for the definition], 
where $\ell$ is measured from the center of the system as shown in Fig.~\ref{fig:Fig3}.
As we have additional contractions coming from the bogolon:
\begin{equation}
\begin{split}
&\langle \eta_{\alpha} {\cal A}_i\rangle=\langle {\cal A}_i\eta_{\alpha}^{\dagger}\rangle=\phi_{i{\alpha}},\\
&\langle \eta_{\alpha} {\cal B}_i\rangle=-\langle {\cal B}_i\eta_{\alpha}^{\dagger}\rangle=-\psi_{i{\alpha}},\\
&\langle \eta_{\alpha}\eta_{\beta}^{\dagger}\rangle=\delta_{{\alpha}\beta} \; ,
\end{split}
\end{equation}
we need to handle the Pfaffian of the matrix of order ($2\ell-2$) instead of the determinant of the matrix of order ($\ell-1$). 

\begin{figure}
\begin{center}
  \includegraphics[scale=1]{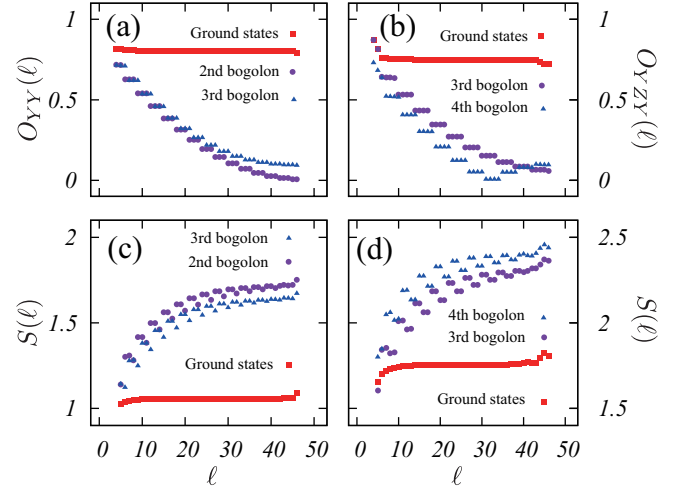} \\
\end{center}
\caption{
(Color Online)
The correlation functions and the block-size ($\ell$) dependence of EE with $N=101$.
(a) The distance dependence of the correlation function $O_{YY}(\ell)$
in the ground state and the single bogolon states $\eta_{2}^{\dagger}|\text{vac}\rangle$ and 
$\eta_{3}^{\dagger}|\text{vac}\rangle$ 
for $(J^{YY}/J^{XZX}, J^{YZY}/J^{XZX})=(2, 0)$.
A triple structure appears for the single bogolon states with a finite energy.
(b) The distance dependence of the correlation function $O_{YZY}(\ell)$
in the ground state and the single bogolon states at $(J^{YY}/J^{XZX}, J^{YZY}/J^{XZX})=(0, 2)$.
A quadruple structure appears for the single bogolon states with a finite energy.
(c) The block-size dependence of the EE $S(\ell)$
in the ground state and the single bogolon states at $(J^{YY}/J^{XZX}, J^{YZY}/J^{XZX})=(2, 0)$.
(d) The block-size dependence of the EE $S(\ell)$
in the ground state and the single bogolon states at $(J^{YY}/J^{XZX}, J^{YZY}/J^{XZX})=(0, 2)$.
}
\label{fig:Fig9}
\end{figure}

With the above setup,
we calculated the correlation functions in the excited states $|\alpha\rangle$ 
of the instantaneous Hamiltonians $H_{1,2}(t=\tau)$.
We show the distance $\ell$ dependence of the correlation functions
$O_{YY}(\ell)_{\alpha}$ for $H_{1}(t=\tau)$ [i.e., $H_{\rm GC}$ with $(J^{YY}/J^{XZX}, J^{YZY}/J^{XZX})=(2, 0)$] 
in Fig.~\ref{fig:Fig9}(a) 
and $O_{YZY}(\ell)_{\alpha}$ for $H_{2}(t=\tau)$ [i.e., $(J^{YY}/J^{XZX}, J^{YZY}/J^{XZX})=(0, 2)$] 
in Fig.~\ref{fig:Fig9}(b).
As mentioned before, the bogolons with zero energy are responsible for the ground-state degeneracy.
Only the first bogolon has zero energy [Figs.~\ref{fig:Fig9}(a) and (c)] in the AF$^{(y)}$ phase,
whereas both the first and second bogolons have zero energy [Figs.~\ref{fig:Fig9}(b) and (d)] in the C* phase.
In fact, the correlation functions in the zero-energy excited states are 
the same as those in the Bogoliubov vacuum,
since these states, together with $|\text{vac}\rangle$, form 
the degenerate set of (topological) ground states.
In Figs.~\ref{fig:Fig9}(a) and (b), we show the correlation functions for the ground states and 
the bogolon excited states with {\em finite} energies.
For the finite-energy bogolon states, the spatially periodic structures are observed in both cases. 
The periods 3 [for $H_{1}(t=\tau)$] and 4 [for $H_{2}(t=\tau)$] may be traced back to the wave lengths 
of the corresponding bogolons.  
 
Next we calculated the EE for the same single bogolon states.
To calculate the EE for excited states, the reduced density matrix is obtained by using Eq.~(\ref{eq:rdm}),
where $\rho$ is defined for the excited state in question.  Then, the EE for excited states is readily obtained  
by applying Eq.~(\ref{eq:ee}).    
We take a block of $\ell$ adjacent sites as the subsystem as shown in Fig.~\ref{fig:Fig3}.
We show the block-size dependence of the EE
at $(J^{YY}/J^{XZX}, J^{YZY}/J^{XZX})=(2, 0)$ in Fig.~\ref{fig:Fig9}(c)
and at $(J^{YY}/J^{XZX}, J^{YZY}/J^{XZX})=(0, 2)$ in Fig.~\ref{fig:Fig9}(d).   
Again, the excitation of a single zero-energy bogolon yields the same EE as 
that for the Bogoliubov vacuum.  
In Figs.~\ref{fig:Fig9}(c) and (d), we show the EE for the ground states $|\text{vac}\rangle$ 
and the two finite-energy bogolon states $|\alpha\rangle=\eta^{\dagger}_{\alpha}|\text{vac}\rangle$ ($\alpha=2,3$ or $3,4$) 
of the two instantaneous Hamiltonians $H_{1}(t=\tau)$ and $H_{2}(t=\tau)$, respectively. 
For the bogolon states with finite energies, spatially periodic (period-3 and 4) structures that are reminiscent of 
what were seen during the sweep after passing the critical points
[i.e., $t > \tau/2$; see Figs.~\ref{fig:Fig4}(d) and \ref{fig:Fig7}(d)] are observed.  
From these results, 
we may conclude that the single-bogolon state $|\alpha\rangle$ with the lowest non-zero energy 
dominates the dynamics in slow sweeps 
and that the periodic structure found in the correlation functions and the EE 
(see Figs.~\ref{fig:Fig4} and \ref{fig:Fig7}) essentially originates from it.
We can observe these structures when the sweep time is larger than typical time at the critical point (i.e. $\tau > N$),
which is a manifestation of the breakdown of adiabaticity.

\begin{figure}
\begin{center}
  \includegraphics[scale=1]{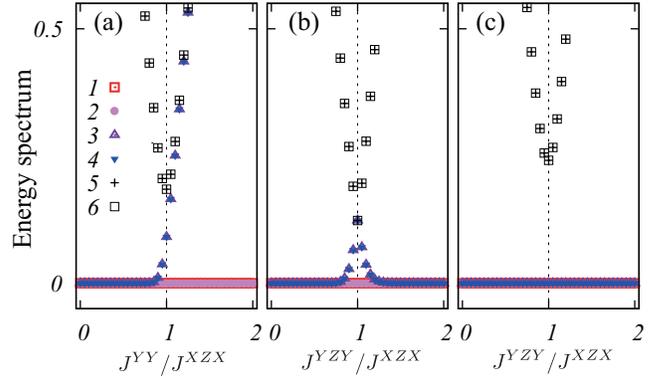} \\
\end{center}
\caption{
(Color Online)
The plot of the lowest-lying six energies of $H_{\rm GC}$ along the path `P1' or `P2' 
(see Fig.~\ref{fig:Fig2}).
(a) Along the path P1 ($J^{YZY}=0$) for $N=101$.   
Along the path P2 ($J^{YY}=0$) for $N=101$ [(b)], and for $N=102$ [(c)].  
The level crossing occurs at $J^{YZY}/J^{XZX}=1$ for $N=101$, 
while an avoided crossing occurs for $N=102$.
}
\label{fig:Fig10}
\end{figure}

This breakdown of adiabaticity is due to the topological properties of the system.
In the C and C* phases, the four ground states
$|{\rm vac}\rangle$, $\eta_1^{\dagger}|{\rm vac}\rangle$, $\eta_2^{\dagger}|{\rm vac}\rangle$, $\eta_2^{\dagger}\eta_1^{\dagger}|{\rm vac}\rangle$
are labeled by the eigenvalues of the parity operators (\ref{eq:parityC}).
In the AF$^{(y)}$ phase, the two ground states $|{\rm vac}\rangle$, $\eta_1^{\dagger}|{\rm vac}\rangle$
are labeled by the eigenvalues of the parity operators (\ref{eq:parity}).
As shown in Sec.~\ref{sec:CAF}, the system undergoes the phase transition between the C and AF$^{(y)}$ phases.
Because these two phases differ in the degree of the ground-state degeneracy, 
some of the degenerate ground states must leave the ground state subspace after the quench and
the breakdown of adiabaticity occurs for the reason discussed in Refs.~\onlinecite{Kells-2014,Hegde-2014}.
In Fig.~\ref{fig:Fig10}(a), we show the energy spectrum of the low-lying states along the path P1 in Fig.~\ref{fig:Fig2}.
Two of the four states, which constitute the degenerate ground states in the C phase, 
are lifted up as we increase the parameter $J^{YY}/J^{XZX}$ across the critical point M1 ($J^{YY}/J^{XZX}=1$).
Since the initial state in the sweep dynamics, i.e., $|{\rm vac}\rangle$, is one of the two states leaving 
the ground-state subspace, the initial state is transferred to an excited state of the final Hamiltonian in the AF$^{(y)}$ side 
even after a slow sweep.

On the other hand, the degree of the degeneracy is four both in the C and C* phases as shown in Sec.~\ref{sec:CC*}.
In this case,
the breakdown of adiabaticity is {\it not} due to the mismatch between the numbers of the ground states as in the case of Sec.~\ref{sec:CAF}.
Rather the system size matters in the following manner.
By using the fermion representation,
the model (\ref{eq:H2}) is rewritten as two independent Kitaev chains [see Figs.~\ref{fig:Fig1}(a) and (c)]:
\begin{equation}
\begin{split}
\label{eq:HoHe}
& H_2(t)= H_{2,\rm odd}(t)+H_{2,\rm even}(t),\\
& H_{2,\rm odd}(t)= J^{XZX}\sum_{i\in {\rm odd}} (-c_i+c_i^{\dagger}) (c_{i+2}+c_{i+2}^{\dagger}) \\
&  \phantom{H_{2,\rm odd}(t)= }
+J^{YZY}(t)\sum_{i\in {\rm odd}} (c_i+c_i^{\dagger}) (-c_{i+2}+c_{i+2}^{\dagger}),\\
& H_{2,\rm even}(t)= J^{XZX}\sum_{i\in {\rm even}} (-c_i+c_i^{\dagger}) (c_{i+2}+c_{i+2}^{\dagger})\\
& \phantom{H_{2,\rm even}(t)= }
+J^{YZY}(t)\sum_{i\in {\rm even}} (c_i+c_i^{\dagger}) (-c_{i+2}+c_{i+2}^{\dagger}).
\end{split}
\end{equation}
The energy-level structure of the chain near the transition is qualitatively different depending on the parity of the system 
size as pointed out in Ref.~\onlinecite{Bermudez-2010};
when the size of the chain is even,
a level crossing between zero-energy level and finite one occurs at the critical point, 
while we observe an avoided level crossing when the size of the chain is odd.  
We show the change in the low-lying energy spectrum along the path P1 
for the system sizes $N=101=51+50$ and $N=102=51+51$
in Figs.~\ref{fig:Fig10}(b) and (c), respectively.
We observe the level crossing for the former case and the avoided level crossing for the latter case.
In the case of $N=101$ studied in Sec.~\ref{sec:CC*},
two of the four degenerate ground states in the C phase are lifted
and two of excited states come down to merge with the ground-state level in the C* side
as we increase the parameter $J^{YZY}/J^{XZX}$ [See Fig.~\ref{fig:Fig10}(b)].  
Since the initial state in the sweep dynamics corresponds to one of the two lifted states,
we observe the breakdown of adiabaticity in the dynamics. 

\subsection{\label{sec:oscillating}Oscillating and splitting structures of the ES}
\begin{figure}
\begin{center}
  \includegraphics[scale=1]{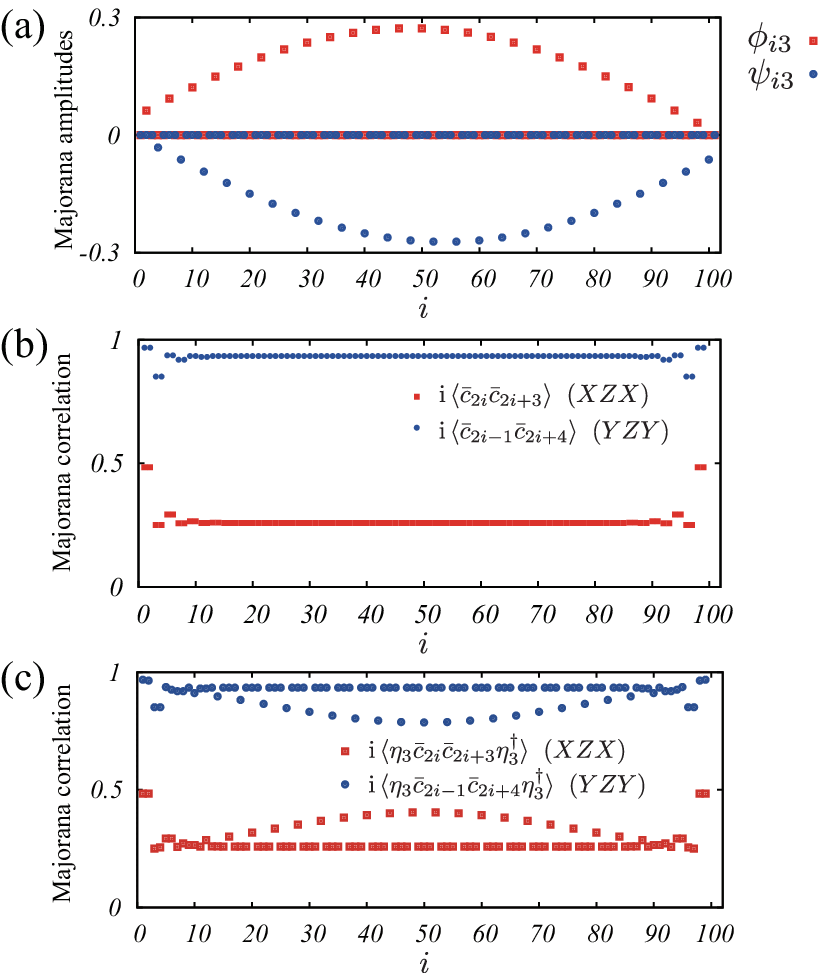} \\
\end{center}
\caption{
(Color Online)
(a) The real-space amplitudes $(\phi,\psi)$ of the third bogolon 
with $N=101$ at $(J^{YY}/J^{XZX}, J^{YZY}/J^{XZX})=(0, 2)$.
(b) The strength of the Majorana correlation in the ground states with $N=101$ at $(J^{YY}/J^{XZX}, J^{YZY}/J^{XZX})=(0, 2)$.
(c) The strength of the Majorana correlation in the excited state $\eta^{\dagger}_{3}|\text{vac}\rangle$ 
with $N=101$ at $(J^{YY}/J^{XZX}, J^{YZY}/J^{XZX})=(0, 2)$.
}
\label{fig:Fig11}
\end{figure}

We discuss the origin of the oscillating and splitting structures of the ES in dynamics
shown in Figs.~\ref{fig:Fig6}(a)-(d) and \ref{fig:Fig8}(a)-(d).
In the previous section, we concluded that the bogolon states with finite energy
play a crucial role in the sweep dynamics across the critical points.
In the following, we concentrate on the case of the sweep dynamics from the C phase to the C* phase (path P2).
Since the third bogolon state has the lowest non-zero energy, we focus on the third bogolon state.
In Fig.~\ref{fig:Fig11}(a), we show the amplitudes $\phi$ and $\psi$ of the third bogolon state 
[see Eq.~\eqref{eq:BogoliubovOperator}].
They are delocalized into the bulk of the system,
whereas the zero-energy modes are localized only at the ends.
In the ground states, the spatial pattern of the Majorana correlation [Figs.~\ref{fig:Fig1}(a), (b), and (c)] determines the topological properties.
Therefore it would be important to know how the excitation affects the spatial pattern.
(See Appendix \ref{sec:dimermodel} for the relationship between the Majorana correlation and the ES.)

\begin{figure}
\begin{center}
  \includegraphics[scale=1]{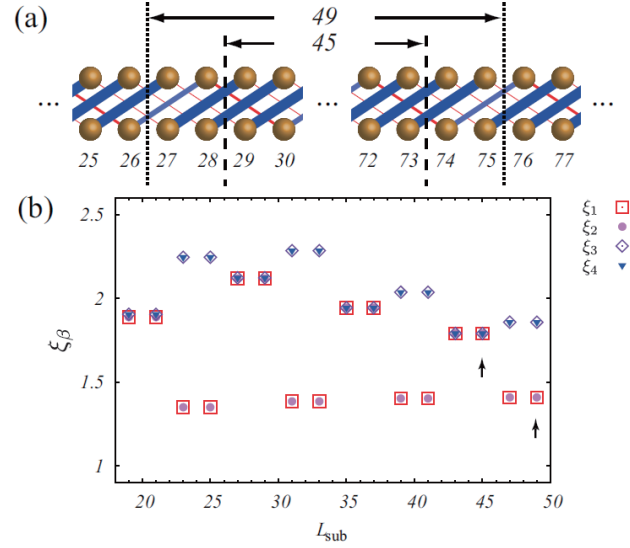} \\
\end{center}
\caption{
(Color Online)
(a) Schematics of the Majorana correlation with $N=101$.
The strength of the Majorana correlation in Fig.~\ref{fig:Fig11}(c) are shown by the thickness of the bonds.
(b) The ES $\{ \xi_{\beta}(L_{\rm sub}) \}$
for the third bogolon state at $(J^{YY}/J^{XZX}, J^{YZY}/J^{XZX})=(0, 2)$ plotted against the size 
$L_{\rm sub}$ of the subsystem (see Fig.~\ref{fig:Fig5} for the arrangement of the subsystem).  
$L_{\rm sub}=45$ and $49$ are marked by the arrows. 
}
\label{fig:Fig12}
\end{figure}

To quantify the Majorana correlation,
we first introduce the correlation function between the $i$-th and the $j$-th Majorana fermions in the vacuum by
\begin{equation}
\label{eq:bondcorrelation}
{\mathrm i}\, \langle\bar{c}_{i}\bar{c}_{j}\rangle.
\end{equation}
For example, the cluster interaction $\sigma_i^x\sigma_{i+1}^z\sigma_{i+2}^x \sim \bar{c}_{2i}\bar{c}_{2i+3}$
contributes to the correlation between the $2i$-th and $(2i+3)$-th Majorana fermions.
In other words, in the cluster phase, ${\mathrm i}\, \langle\bar{c}_{2i}\bar{c}_{2i+3}\rangle$ is finite.
In the analysis of the excited states,
we calculate the Majorana correlation (\ref{eq:bondcorrelation}) 
in the excited states $|\alpha\rangle$,
that is, ${\mathrm i}\, \langle\eta_\alpha \bar{c}_{2i}\bar{c}_{2i+3} \eta_\alpha^\dagger\rangle$.
Since we here consider only the third bogolon state, we set $\alpha = 3$.
In Figs.~\ref{fig:Fig11}(b), (c), we show the spatial dependence of the correlation 
between the $2i$-th and $(2i+3)$-th Majorana fermions (`$XZX$ bond') 
and that between the $(2i-1)$-th and $(2i+4)$-th Majorana fermions (`$YZY$ bond').
The former (latter) detects the correlation characteristic of the cluster interaction ($J^{XZX}$) 
[the dual cluster interaction ($J^{YZY}$)].
The Majorana correlation in the ground states 
barely depends on the position of the bond except near the ends as is seen in Fig.~\ref{fig:Fig11}(b).
For the third bogolon state $\eta^{\dagger}_{3}|\text{vac}\rangle$, on the other hand, 
they exhibit peculiar structures as shown in Fig.~\ref{fig:Fig11}(c);
the Majorana correlations are significantly affected by the bogolon every four bonds,
while they are almost intact at the other bonds [see Fig.~\ref{fig:Fig12}(a)].   
This suggests that we may explain the dynamics of the ES in terms of the Majorana correlation.
To substantiate this, we represent the strength of the Majorana correlation 
by the thickness of the bonds in Fig.~\ref{fig:Fig12}(a).
When the $YZY$ bond between the $(2i-1)$-th and $(2i+4)$-th Majorana fermions becomes weaker,
the $XZX$ bond between the $(2i-4)$-th and $(2i-1)$-th Majorana fermions becomes stronger.

Next we discuss the relationship between the Majorana correlation and the ES defined for an excited state.
Because the ES in the ground state is believed to reflect the emergent degrees of freedom 
at the boundaries,\cite{Li-2008}  
we expect that some information in the excited states can be obtained from the ES  
as well as the EE shown in the previous section.
As in the previous section, we consider the reduced density matrix obtained by Eq.~(\ref{eq:rdm}),
where $\rho=|\Psi\rangle\langle \Psi|$ is the density matrix of the third bogolon state 
$|\Psi\rangle=\eta^{\dagger}_{3}|\text{vac}\rangle$.
Substituting the eigenvalues of $\rho$ in Eq.~(\ref{eq:es}), we can obtain the ES for the excited state in question.
Here we cut the system as shown in Fig.~\ref{fig:Fig5}.
We calculated the block-size ($L_{\rm sub}$) dependence of the ES for the third bogolon state.
The lowest four entanglement levels are shown in Fig.~\ref{fig:Fig12}(b).
The degeneracy depends on the length of the subsystem. 

Let us focus on the specific cases: $L_{\rm sub}=45$ and 49.
First, when $L_{\rm sub} = 45$, the boundaries between two subsystems are indicated by the dashed lines in Fig.~\ref{fig:Fig12}(a).
In this case, we cut four thick $YZY$ (blue) bonds which are not affected at all by the third bogolon, 
and as a consequence we have four unpaired Majorana fermions at the ends.  
Because they form two fermionic excitations with zero energy, 
the lowest level of the ES shows four-fold quasi-degeneracy shown in Fig.~\ref{fig:Fig12}(b) 
(the level indicated by the left arrow). 

Second, when $L_{\rm sub}=49$, the boundaries between two subsystems are shown 
by the dotted lines in Fig.~\ref{fig:Fig12}(a).
Now we cut two thick and two thin $YZY$ bonds.
In this case, only two unpaired Majorana fermions appear at the cut ends and 
the other Majorana fermions are disturbed by the third bogolon.
The two unpaired Majorana fermions contribute to the double degeneracy of the ES in Fig.~\ref{fig:Fig12}(b) 
(see the level shown by the right arrow).  
Therefore the ES for the excited states reflects the strength of the Majorana correlation 
when the model is quadratic in the Majorana fermions.

Finally we explain the peculiar time evolution of the ES observed  in Sec.~\ref{sec:CC*} (see Fig.~\ref{fig:Fig8}).
Here we set $L_{\rm sub}=49$ in the calculation.
Because the dynamics after passing the critical point is dominated by the third bogolon for slow enough sweeps,
we may expect that the dynamical behavior of the ES may be captured essentially by the third bogolon state.
As we have seen in Fig.~\ref{fig:Fig12}(b),
the number of degeneracy of the ES is two.
This is the origin of the splitting of the ES for the slower sweep.
On the other hand, for faster sweeps,
the excitations whose energies are higher than that of the third bogolon also contribute to the dynamics
after passing the critical point.  
The final state at $t=\tau$ is a superposition of those excited states,
which was already studied in Ref.~\onlinecite{Ohta-2015}.
Because the state is not the eigenstate of the instantaneous Hamiltonian at time $t >\tau/2$, 
the expectation values of the Majorana correlation in the state oscillate in time.
Therefore the ES oscillates in time for a faster sweep.
A similar argument applies to the oscillating structure in Fig.~\ref{fig:Fig6}.

\section{\label{sec:Conclusion}Conclusion}
We have studied the ground-state phase diagram and dynamics of the cluster model 
in one dimension with several competing interactions.
First, we have determined boundaries among several quantum phases of the model by the energy gap
and identified the universality classes of the critical points using several CFT techniques.    
The phases are then characterized by the winding number and the ES,
which reflect the number of the Majorana zero modes.  
A lot of phases appear as a result of the competition among several Majorana interactions. 

Second, we have investigated dynamical properties during two types of interaction sweep through the critical points 
which separate two topological phases: 
the C phase to the AF$^{(y)}$ phase and the C phase to the C* phase.
After slow sweeps across the critical points, spatially periodic structures have been observed 
both in the correlation functions and in the EE 
and oscillating and splitting structures have been found in the ES.  
This implies that even for slow enough sweeps, the ground state of one phase evolves into 
the final state which is no longer the ground state of the instantaneous Hamiltonian.  
This breakdown of adiabaticity is due to the fact that the degenerate ground states are labeled by the eigenvalues 
of the fermionic parity which determine the topological properties of the system.  
Unlike the usual sweep dynamics across critical points,\cite{Caneva-2007,Dziarmaga-2006,Das-2008,Suzuki-2011}
the dynamical behavior observed here is not characterized by the critical exponents but the fermion-number
parity of the initial state.
Finally, we have tried to reproduce the similar structures by using the low-lying excited states 
and verified that the structures come from a single bogolon excited near the critical points.
In addition, we have found that the ES reflects the strength of the Majorana correlation even for the excited states.
\section*{Acknowledgements}
This work was supported by JSPS KAKENHI Grant Numbers 15K17720, 15H03699, 25420698 (S.T.),
25800228, 25220711 (I.D.),
15K05211 (K.T.).
S.T. was also supported by Waseda University Grant for Special Research Projects (Project number:2015B-514).
I.D. was also supported by the "Topological Materials Science"(No.15H05855) KAKENHI on innovative Areas from MEXT of Japan.
The computations in the present work were performed on super computers at Yukawa Institute for Theoretical Physics, Kyoto University,
and Institute for Solid State Physics, The University of Tokyo.

\appendix
\section{Operator content and universality class}
\label{sec:operator-content}
In this appendix, we determine the operator content of the conformal field theories describing 
the critical points. 
To be specific, we take the periodic boundary condition and 
consider the multicritical point M1: $(J^{YY}/J^{XZX},J^{YZY}/J^{XZX})=(-1,0)$ 
where the gapless $k$-linear branches exist at $k=0, \pm 2\pi/3$.  
Then, the low-energy physics may be described by the effective Hamiltonian 
\begin{equation}
\begin{split}
& H_{\text{GC}} - E_{\text{ZP}} \\
& \approx \sum_{|k| \leq \Lambda} v_{\text{s}} k \, \eta^{\dagger}_{k-\frac{2\pi}{3}}\eta_{k-\frac{2\pi}{3}} 
+ \sum_{|k| \leq \Lambda} v_{\text{s}} k \, \eta^{\dagger}_{k}\eta_{k}
+ \sum_{|k| \leq \Lambda} v_{\text{s}} k \, \eta^{\dagger}_{k+\frac{2\pi}{3}}\eta_{k+\frac{2\pi}{3}}  \\
& \equiv \sum_{|k| \leq \Lambda} v_{\text{s}} k \, \eta(1)^{\dagger}_{k}\eta(1)_{k} 
+ \sum_{|k| \leq \Lambda} v_{\text{s}} k \, \eta(2)^{\dagger}_{k}\eta(2)_{k}
+ \sum_{|k| \leq \Lambda} v_{\text{s}} k \, \eta(3)^{\dagger}_{k}\eta(3)_{k} \; ,
\end{split}
\end{equation}
where the common ``light velocity'' $v_{\text{s}}=6 J^{XZX}$ and $E_{\text{ZP}}$ denotes 
the regularized zero-point energy.  
The summation over $k$ should be cut off at $\Lambda$ which gives the bound for the linearization 
of the spectrum.  
Due to the non-local nature of the Jordan--Wigner transformation \eqref{eqn:JW-tr}, 
the boundary condition (or, the allowed values of momentum $k$) 
for the fermion depends explicitly on the total fermion number $F$:
\begin{equation}
k = 
\begin{cases}
\frac{2\pi}{N}(j+1/2) \;\; (j=0,\ldots, N-1) & \quad \text{when $F$ is even} \\
\frac{2\pi}{N} j  \;\; (j=0,\ldots, N-1) & \quad \text{when $F$ is odd} \; .
\end{cases}
\label{eqn:JW-PBC}
\end{equation}
When the system size $N$ is an integer-multiple of $3$, the above moding carries over to 
the individual branches $\eta(i)$ ($i=1,2,3$)\cite{comment-CFT}
and the zero-point energy is given by
\begin{equation}
E_{\text{ZP}}  =
\begin{cases}
N \epsilon_{\infty} - \frac{\pi v_{\text{s}}}{6N} \frac{3}{2}  & F=\text{even} \\
N \epsilon_{\infty} - \frac{\pi v_{\text{s}}}{6N} \frac{3}{2} + \frac{2\pi v_{\text{s}}}{N} \frac{3}{8}  & F=\text{odd}  
\end{cases} 
\end{equation}  
up to $O(1/N)$.  

In order to take into account the fermion-number dependence of the boundary condition, 
it is convenient to introduce the following projection operator which is written in terms of 
the fermion numbers $\{ F^{(i)}_{\text{R/L}} \}$ of the individual Majorana branches:
\begin{equation}
\begin{split}
& \mathcal{P}_{\pm} \equiv \frac{1}{2} \left[ 1 \pm (-1)^{F} \right] \\
& F = F^{(1)}_{\text{L}} + F^{(1)}_{\text{R}} + F^{(2)}_{\text{L}} 
+ F^{(2)}_{\text{R}} + F^{(3)}_{\text{L}} + F^{(3)}_{\text{R}}  \; ,
\end{split}
\end{equation}
where $\mathcal{P}_{+}$ ($\mathcal{P}_{-}$) must be used with anti-periodic (periodic) moding. 
As $F^{(i)}_{\text{R}} = \sum_{k>0} \eta(i)^{\dagger}_{k}\eta(i)_{k}$, 
$F^{(i)}_{\text{L}} = \sum_{k>0} \eta(i)^{\dagger}_{-k}\eta(i)_{-k}$, 
the partition function for the even-$F$ sector (at temperature $T$) is calculated as
\begin{equation}
\begin{split}
& Z_{F\text{-even}} \\
& = q^{-\frac{3}{48}} \bar{q}^{-\frac{3}{48}} \text{Tr}\, 
\mathcal{P}_{+} \, 
q^{\sum_{i=1}^{3} \sum_{n_i >0} n_i \eta(i)^{\dagger}_{n_i}\eta(i)_{n_i}}
\bar{q}^{\sum_{n_i} \sum_{n_i >0} n_i \eta(i)^{\dagger}_{-n_i}\eta(i)_{-n_i}} \\
&= \frac{1}{2}(q\bar{q})^{-\frac{3}{48}}  \Biggl\{ 
\left(\text{Tr}\,  q^{\sum_{n>0} n \eta^{\dagger}_{n}\eta_{n}}\right)^{3}
\left( \text{Tr}\, \bar{q}^{\sum_{n>0} n \eta^{\dagger}_{-n}\eta_{-n}} \right)^{3} \\
& \phantom{=}
+ \left( \text{Tr}\,  (-1)^{F_{\text{L}}}q^{\sum_{n>0} n \eta^{\dagger}_{n}\eta_{n}} \right)^{3}
\left( \text{Tr}\, (-1)^{F_{\text{R}}}\bar{q}^{\sum_{n>0} n \eta^{\dagger}_{-n}\eta_{-n}}  \right)^{3} 
\Biggr\} \\
&= \frac{1}{2}(q\bar{q})^{-\frac{3}{48}}  \left\{ 
\left[ \prod_{n=0}^{\infty} (1+q^{n+\frac{1}{2}}) 
\prod_{\bar{n}=0}^{\infty} (1+\bar{q}^{\bar{n}+\frac{1}{2}}) \right]^{3}  \right. \\
& \phantom{=} \left.
+ \left[ \prod_{n=0}^{\infty} (1-q^{n+\frac{1}{2}})\prod_{\bar{n}=0}^{\infty} (1-\bar{q}^{\bar{n}+\frac{1}{2}}) \right]^{3} 
\right\}  \\
&= \frac{1}{2} \left( 
\left\vert \frac{\vartheta_{3}(q)}{\eta(q)} \right\vert^{3} 
+ \left\vert \frac{\vartheta_{4}(q)}{\eta(q)} \right\vert^{3} \right) 
 \; ,
\end{split}
\end{equation}
where 
\begin{equation} 
q = \bar{q} =\exp \left( - \frac{2\pi v_{\text{s}}}{TN} \right) 
\end{equation}
and the Boltzmann constant is set to unity.
The $\vartheta_{i}(q)$ are the Jacobi theta functions\cite{Gradshteyn-2000} 
$\vartheta_{i}(q)\equiv \vartheta_{i}(0, q^{\frac{1}{2}})$ and 
\begin{equation}
\eta(q) = q^{\frac{1}{24}} \prod_{n=1}^{\infty} (1-q^{n}) \; .
\end{equation}
Similarly, the partition function for the subspace with odd fermion numbers $Z_{F\text{-odd}}$ is given by
\begin{equation}
\begin{split}
& Z_{F\text{-odd}} \\
&= q^{-\frac{3}{48}} \bar{q}^{-\frac{3}{48}} \text{Tr}\, 
\mathcal{P}_{-}  \, 
q^{\sum_{i}\sum_{n_i \geq 0} \left[ n_i \eta(i)^{\dagger}_{n_i}\eta(i)_{n_i}+\frac{1}{16} \right]}
\bar{q}^{\sum_{i} \sum_{n_i >0} \left[ n_i \eta(i)^{\dagger}_{-n_i}\eta(i)_{-n_i}+\frac{1}{16} \right] } \\
&= \frac{1}{2}(q\bar{q})^{\frac{3}{24}}  \left\{ 
\left[
\prod_{n=0}^{\infty} (1+q^{n})\prod_{\bar{n}=1}^{\infty} (1+\bar{q}^{\bar{n}}) 
\right]^{3} \right.  \\
& \phantom{=\frac{1}{2}(q\bar{q})^{\frac{3}{24}} \qquad } 
+\left.  \left[ \prod_{n=0}^{\infty} (1-q^{n})\prod_{\bar{n}=1}^{\infty} (1-\bar{q}^{\bar{n}}) 
\right]^{3} 
\right\}  \\
&= \frac{1}{2}  \left\vert \frac{\vartheta_{2}(q)}{\eta(q)} \right\vert^{3}     \; .
\end{split}
\end{equation}
In deriving the above, we have used the fact that the zero mode $n=0$ is occupied by either $\eta(i)_{\text{R}}$ or 
$\eta(i)_{\text{L}}$ [here $\eta(i)_{\text{R}}$].  
Therefore, the full (low-energy) partition function at the multicritical point M1 reads as
\begin{equation}
Z_{\text{M1}} = Z_{F\text{-even}} + Z_{F\text{-odd}}  
= \frac{1}{2} \left( 
\left\vert \frac{\vartheta_{3}(q)}{\eta(q)} \right\vert^{3} 
+ \left\vert \frac{\vartheta_{4}(q)}{\eta(q)} \right\vert^{3} 
+ \left\vert \frac{\vartheta_{2}(q)}{\eta(q)} \right\vert^{3} 
\right)  \; ,
\end{equation}
which is {\em different} from that of three decoupled Ising models:
\begin{equation}
(Z_{\text{Ising}})^{3} = \left\{
\frac{1}{2} \left( 
\left\vert \frac{\vartheta_{3}(q)}{\eta(q)} \right\vert 
+ \left\vert \frac{\vartheta_{4}(q)}{\eta(q)} \right\vert 
+ \left\vert \frac{\vartheta_{2}(q)}{\eta(q)} \right\vert
\right) \right\}^{3} \; .
\end{equation}
In fact, we can show that $Z_{\text{M1}}$ is equivalent to the partition function 
of the level-1 SO(3) WZW model.\cite{Witten-1984,Knizhnik-1984}   
To see this, it is convenient to expand $q^{\frac{3}{48}}\bar{q}^{\frac{3}{48}}Z_{\text{M1}}$ 
in a power series of $q$ and $\bar{q}$:
\begin{equation}
q^{\frac{3}{48}}\bar{q}^{\frac{3}{48}}Z_{\text{M1}} = 
1 + 4 q^{\frac{3}{16}}\bar{q}^{\frac{3}{16}} + 9 q^{\frac{1}{2}}\bar{q}^{\frac{1}{2}} + 3q + 3\bar{q} + \cdots
\end{equation}
For instance, the coefficient 3 of $q$ ($\bar{q}$) in $q^{\frac{3}{48}}\bar{q}^{\frac{3}{48}}Z_{\text{M1}}$ 
coincides with the number of the left (right) SO(3) currents.  
Similarly, the coefficient 4 (9) of $(q\bar{q})^{\frac{3}{16}}$ [$(q\bar{q})^{\frac{1}{2}}$] comes from 
the number of the WZW primary fields with $(h,\bar{h})=\left(\frac{3}{16},\frac{3}{16} \right)$ 
[$(h,\bar{h})=\left(\frac{1}{2},\frac{1}{2} \right)$] transforming under the spinor (vector) representation 
of SO(3).\cite{CFT-yellow-book}  
Note that the non-trivial boundary condition \eqref{eqn:JW-PBC} has led us to summing over 
all possible fermionic boundary conditions and reproduces the correct partition function of the WZW model.  
We can follow similar steps to derive the partition function corresponding to the multicritical point M2:
\begin{equation}
Z_{\text{M2}} 
= \frac{1}{2} \left( 
\left\vert \frac{\vartheta_{3}(q)}{\eta(q)} \right\vert^{4} 
+ \left\vert \frac{\vartheta_{4}(q)}{\eta(q)} \right\vert^{4} 
+ \left\vert \frac{\vartheta_{2}(q)}{\eta(q)} \right\vert^{4} 
\right)  \; ,
\end{equation}
which implies that the critical point M2 is described by the level-1 SO(4) WZW model.  
\section{\label{sec:dimermodel}Dimer model}
\begin{figure}
\begin{center}
  \includegraphics[scale=1]{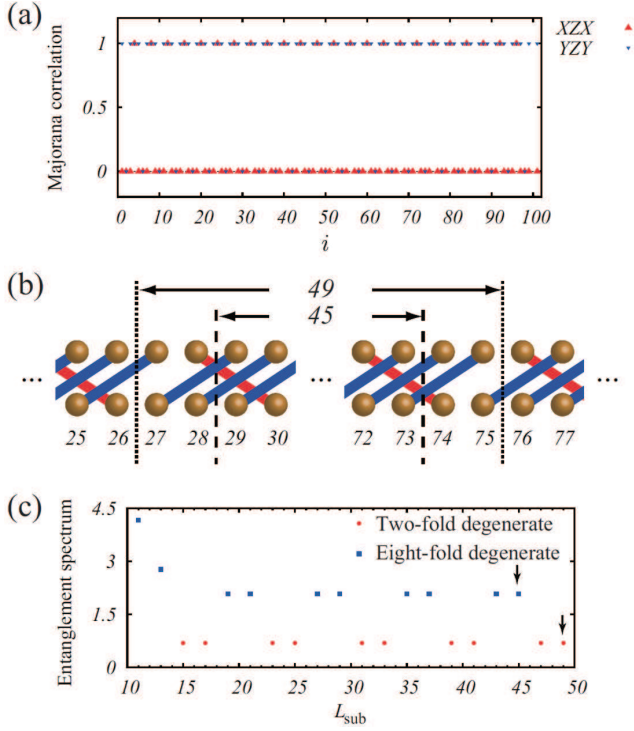} \\
\end{center}
\caption{
(Color Online)
(a) The strength of the Majorana correlation for the ground state of the model (\ref{eq:HD}) with $N=101$.
(b) The picture of the Majorana correlation with $N=101$.
The strength of the Majorana correlation in Fig.~\ref{fig:Fig13}(a) is shown by the bonds.
(c) The block-size dependence of the ES $\xi_{\beta}(L_{\rm sub})$ for the vacuum with $N=101$.
}
\label{fig:Fig13}
\end{figure}

In Sec.~\ref{sec:oscillating},
we investigated the relationship between the Majorana correlation defined by Eq.~(\ref{eq:bondcorrelation}) and the ES.
To confirm the relationship between the Majorana correlation and the ES,
we calculate the ES of a model whose Majorana correlation is obvious.
We consider a model defined by
\begin{equation}
\label{eq:HD}
H_{\rm D}=\sum_{i=1}^{N}(-J^{XZX}_i\sigma_i^x\sigma_{i+1}^z\sigma_{i+2}^x+J^{YZY}_{i}\sigma_i^y\sigma_{i+1}^z\sigma_{i+2}^y),
\end{equation}
where
\begin{align}
J_i^{XZX}=
\begin{cases}
0 \,\,\, {\rm if}\, i\neq 0\, ({\rm mod} 4)\\
2 \,\,\, {\rm if}\, i= 0 \,({\rm mod} 4)\\
\end{cases}
J_i^{YZY}=
\begin{cases}
1 \,\,\, {\rm if}\, i\neq 2\, ({\rm mod} 4)\\
0 \,\,\, {\rm if}\, i= 2 \,({\rm mod} 4)\\
\end{cases}.
\end{align}
Although degenerate ground states exist as well as the generalized cluster model we considered,
we here focus on the Bogoliubov vacuum state.
In Fig.~\ref{fig:Fig13}(a), we show the Majorana correlation
between the $2i$-th and $(2i+3)$-th Majorana fermions ($XZX$) and the $(2i-1)$-th and $(2i+4)$-th Majorana fermions ($YZY$).
The Majorana correlation in the vacuum state is depicted in Fig.~\ref{fig:Fig13}(b).
Here a Majorana fermion interacts with at most one Majorana fermion.
The block-size dependence of the ES in the vacuum state is shown in Fig.~\ref{fig:Fig13}(c).
When $L_{\rm sub}=45$, for example, the boundaries between two subsystems are indicated by the dashed lines in Fig.~\ref{fig:Fig13}(b).
We cut four $YZY$ bonds and two $XZX$ bonds are cut.
Therefore six unpaired Majorana fermions exist at the cut ends.
Because they form three fermionic excitations with zero energy,
the lowest level of the ES shows the eight-fold degeneracy in Fig.~\ref{fig:Fig13}(c).
On the other hand, when $L_{\rm sub}=49$,
the boundaries between two subsystems are indicated by the dotted lines in Fig.~\ref{fig:Fig13}(b).
We cut two $YZY$ bonds (see Fig.~\ref{fig:Fig13}(c)).
In this case, we have two unpaired Majorana fermions at the cut ends,
which contribute to the two-fold degeneracy of the ES in Fig.~\ref{fig:Fig13}(c).
We thus see that the number of unpaired Majorana fermions at the cut ends in the dimer model (\ref{eq:HD})
is reflected in the degeneracy structure of the ES,
as in the case of the Su--Schrieffer--Heeger model.\cite{Sirker-2014}

\begin{thebibliography}{99}
\bibitem{comment-CFT}Unless this condition is met, the continuum limit is ill-defined and we cannot construct the vacuum of the corresponding CFT.
\end{thebibliography}
\end{document}